\documentclass[fleqn,usenatbib,usedcolumn]{mnras}

\usepackage[british]{babel}
\usepackage{newtxtext}
\usepackage[slantedGreek]{newtxmath}
\let\la=\lesssim

\usepackage[T1]{fontenc}
\usepackage{ae,aecompl}

\usepackage{graphicx}
\usepackage{amssymb}
\usepackage{amsmath}
\usepackage{siunitx} 
\usepackage{multicol}
\usepackage{multirow}
\usepackage{float}
\usepackage[caption = false]{subfig}

\usepackage{booktabs,caption}
\usepackage[flushleft]{threeparttable}
\newcommand{\angstrom}{\mbox{\normalfont\AA}}
\usepackage[mathscr]{euscript}

\begin{document}

\label{firstpage}
\pagerange{\pageref{firstpage}--\pageref{lastpage}}

\title[The thermal SZ effect in quasar environments]{Quantifying the Thermal Sunyaev-Zel'dovich Effect and Excess Millimeter Emission in Quasar Environments}

\author[K. R. Hall et al.]{Kirsten R. Hall,$^{1}$\thanks{Email: khall33@jhu.edu}
Nadia L. Zakamska,$^{1}$
Graeme E. Addison,$^{1}$
Nicholas Battaglia,$^{2}$
\newauthor
Devin Crichton,$^{3}$ 
Mark Devlin,$^{4}$
Joanna Dunkley,$^{5}$
Megan Gralla,$^{6}$
J. Colin Hill,$^{7,8}$
\newauthor
Matt Hilton,$^{3}$
Johannes Hubmayr,$^{9}$
John P. Hughes,$^{10}$
Kevin M. Huffenberger,$^{11}$
\newauthor
Arthur Kosowsky,$^{12}$
Tobias A. Marriage,$^{1}$
Lo{\"i}c Maurin,$^{13}$
Kavilan Moodley,$^{3}$
\newauthor
Michael D. Niemack,$^{2}$
Lyman A. Page,$^{5}$
Bruce Partridge,$^{14}$
Rolando D{\"u}nner Planella,$^{13}$
\newauthor
Alessandro Schillaci,$^{15}$
Crist{\'o}bal Sif{\'o}n,$^{16, 17}$
Suzanne T. Staggs,$^{5}$
Edward J. Wollack$^{18}$ 
\newauthor 
and Zhilei Xu$^{4}$
\\
\\
$^{1}$ Department of Physics and Astronomy, Johns Hopkins University, Baltimore, MD 21218, USA \\
$^{2}$ Department of Physics, Cornell University, Ithaca, NY 14853, USA \\
$^{3}$ Astrophysics and Cosmology Research Unit, School of Mathematics, Statistics and Computer Science, University of KwaZulu--Natal, 
\\ \,\,\,  Durban 4041, South Africa \\
$^{4}$ Department of Physics and Astronomy, University of Pennsylvania, Philadelphia, PA 19104-6396, USA \\
$^{5}$ Joseph Henry Laboratories of Physics, Jadwin Hall, Princeton University, Princeton, NJ 08544, USA \\
$^{6}$ Department of Astronomy and Steward Observatory, University of Arizona, Tucson, AZ 85721, USA \\
$^{7}$ School of Natural Sciences, Institute for Advanced Study, Princeton, NJ, USA 08540 \\ 
$^{8}$ Center for Computational Astrophysics, Flatiron Institute, New York, NY, USA 10003 \\
$^{9}$ National Institute of Standards and Technology, Quantum Sensors Group, Boulder, CO \\
$^{10}$ Department of Physics and Astronomy, Rutgers University, 136 Frelinghuysen Road, Piscataway, NJ 08854-8019, USA \\
$^{11}$ Department of Physics, Florida State 
University, Tallahassee, FL 32306, USA \\
$^{12}$ Department of Physics and Astronomy, University of Pittsburgh, Pittsburgh, PA 15260, USA \\
$^{13}$ Instituto de Astrof\'isica and Centro de Astro-Ingenier\'ia, Facultad de F\'isica, Pontificia Universidad Cat\'olica de Chile, Av.  \\
\,\,\, Vicu\~na Mackenna 4860, 7820436 Macul, Santiago, Chile \\
$^{14}$ Department of Physics and Astronomy, Haverford College, Haverford, PA 19041, USA \\
$^{15}$ California Institute of Technology, Division of Physics, Mathematics and Astronomy, Pasadena, CA 91125, USA \\
$^{16}$ Department of Astrophysical Sciences, Princeton University, Princeton, NJ 08544, USA \\
$^{17}$ Instituto de F\`isica, Pontificia Universidad Cat\`olica de Valpara\`iso, Casilla 4059, Valpara\`iso, Chile \\
$^{18}$ NASA/Goddard Space Flight Center, Greenbelt, MD 20771, USA 
}

\date{}

\pubyear{2019}

\maketitle

\begin{abstract}
In this paper we probe the hot, post-shock gas component of quasar-driven winds through the thermal Sunyaev-Zel'dovich (tSZ) effect. 
Combining datasets from the Atacama Cosmology Telescope, the \textit{Herschel} Space Observatory, and the Very Large Array, we measure average spectral energy distributions (SEDs) of 109,829 optically-selected, radio quiet quasars from 1.4~GHz to 3000~GHz in six redshift bins between $0.3<z<3.5$.
We model the emission components in the radio and far-infrared, plus a spectral distortion from the tSZ effect. 
At $z>1.91$, we measure the tSZ effect at $3.8\sigma$ significance with an amplitude corresponding to a total thermal energy of $3.1\times10^{60}$~ergs.
If this energy is due to virialized gas, then our measurement implies quasar host halo masses are $\sim6\times10^{12}~h^{-1}$M$_\odot$.
Alternatively, if the host dark matter halo masses are $\sim2\times10^{12}~h^{-1}$M$_\odot$ as some measurements suggest, then we measure a $>$90 per cent excess in the thermal energy over that expected due to virialization.
If the measured SZ effect is primarily due to hot bubbles from quasar-driven winds, we find that $(5^{+1.2}_{-1.3}$) per cent of the quasar bolometric luminosity couples to the intergalactic medium over a fiducial quasar lifetime of 100 Myr. 
An additional source of tSZ may be correlated structure, and further work is required to separate the contributions.
At $z\leq1.91$, we detect emission at 95 and 148~GHz that is in excess of thermal dust and optically thin synchrotron emission. We investigate potential sources of this excess emission, finding that CO line emission and an additional optically thick synchrotron component are the most viable candidates.
\end{abstract}

\begin{keywords}
galaxies: evolution -- galaxies: active - galaxies: intergalactic medium - quasars: general
\end{keywords} 

\section{Introduction}
The accretion of material onto a supermassive black hole in the nucleus of a galaxy is one of the most energetic phenomena in our universe. 
Quasars are the most powerful class of such Active Galactic Nuclei (AGN) with bolometric luminosities $L_{\rm bol} > 10^{45}$~erg~s$^{-1}$, sometimes outshining all the stars in their host galaxies. 
It has been evident for over three decades that quasar activity is strongly linked to the evolution of the host galaxy.
The evidence driving this conclusion lies in the correlation between the masses of supermassive black holes and their host galaxy bulge mass, luminosity, and velocity dispersion  \citep[e.g.,][]{korm95}, indicating that growth of the super massive black hole is linked to the gas reservoir from which stars form \citep{hopk06}. 
Further  evidence for the connection between quasars and galaxy evolution is the similarity between the observed star formation rate and quasar luminosity density throughout cosmic time (\citealt{boyl98}; \citealt{hopk08}). 

Feedback from accreting black holes has become a key element in modelling galaxy evolution (\citealt{tabo93}; \citealt{silk98}; \citealt{spri05}). 
Quasar feedback is routinely invoked in galaxy formation models to quench star formation and to explain the steep decline of the bright end of the galaxy luminosity function \citep{crot06} as well as to reheat the intracluster medium (\citealt{rawl04}; \citealt{scan04}). 
Numerical simulations of galaxy formation suggest that quasar feedback plays the dominant role in heating the circumgalactic medium (CGM) on scales of hundreds of kpc \citep{pill18}. 

Supermassive black holes are natural candidates for these feedback processes because putting even a small fraction of the accretion energy into an outflow wind is in principle sufficient to liberate the galaxy-scale bound gas from the galaxy's gravitational potential. Although the effects of quasars on their host galaxies and their circumgalactic environment are now thought to be among the major factors in galaxy formation theory, detecting and quantifying this phenomenon from direct observations has proved difficult. 
In the last decade the first observations of galaxy-wide quasar-driven winds have been finally made using optical ionized-gas diagnostics (e.g. \citealt{nesv08, moe09, liu13b, harr14}), and molecular-gas diagnostics (e.g., \citealt{fisc10, veil13a, rupk13, sun14}). 

Despite these observational developments, the amount of energy that the active nucleus is capable of injecting into the extended interstellar medium of the host galaxy remains controversial, and there is no consensus on how quasar winds are launched or how they become coupled to the host's interstellar medium.
The greatest observational challenge to quantifying quasar feedback is the multi-phase nature of quasar-driven winds.
The relatively dense warm ionized emission-line component of winds coexists with a colder, denser neutral and even a molecular component and a more diffuse far-UV/soft X-ray emitting plasma at $10^5-10^6$~K, likely originating on the surfaces of shocked or photo-ionized clouds \citep{gree14a}. Different wind components are observed via different diagnostics across the electromagnetic spectrum.

The most mysterious component of quasar feedback is the lowest--density, hot, volume-filling gas.
This component is predicted by theoretical models of quasar winds (\citealt{fauc12b, zubo12, nims15}): as the winds propagate through the host galaxy and CGM, they leave behind post-shock gas which is too diffuse and too hot ($T > 10^8$ K) to be detectable currently using any emission diagnostics \citep{nims15}. 
With advances in arcminute-resolution millimeter-wavelength telescopes, recent studies described below have been probing this elusive, hot plasma component of quasar winds using the Sunyaev-Zel'dovich effect \citep{suny70}. 

The SZ effect is a distortion of the cosmic microwave background (CMB) spectrum at millimeter wavelengths that occurs when the relatively low-energy CMB photons inverse-Compton scatter off hot ionized gas, such as the volume-filling gas phase of quasar-driven winds. The presence of the SZ effect in galaxies with feedback mechanisms has been predicted by both analytic estimates (\citealt{nata99, yama99, plat02, pfro05, chat07}) and by simulations (\citealt{chat08, scan08, prok10, prok12, cen15}). The magnitude of the total volume-integrated SZ effect is directly proportional to the total thermal energy of the intervening gas. Over the lifetime of a quasar, the energetic output from the winds has the potential to generate a hot plasma out to tens of kiloparsecs, and thus, it is extensive enough to intervene with CMB photons with significant optical depth for CMB photon scattering.
Because the SZ distortion depends linearly on both optical depth and gas temperature, it provides a probe of the thermalized quasar output energy that is not detectable by other methods.
Furthermore, the SZ effect is not subject to surface-brightness dimming as a function of redshift, enabling the study of the low-density, ionized gas in high-$z$ systems.

While detection of the SZ effect in large galaxy clusters dates to the 1990's, the past decade has seen the rise of large blind surveys that can detect the SZ effect from clusters with halo masses $10^{14}-10^{15}~\mathrm{M}_\odot$;  (e.g., \citealt{blee15,plan15b,hilt18}). AGN feedback in these systems is estimated at $\sim$10$^{45}$~erg/s, or $1\times10^{60}$ ergs over a characteristic $100$~Myr timescale \citep[e.g.,][]{cava10}. 
This level of energy injection is dwarfed by the $\ge 2\times10^{62}$~ergs of thermal energy characteristic of virialized gas in galaxy clusters. 
Therefore, disentangling the feedback processes embedded in galaxy clusters through the integrated SZ signal is not yet practical. 
However, these same SZ surveys can be used to study lower mass systems through averaging (stacking) the millimeter-wave data at the locations of the known clusters or sources.
This has been done for massive elliptical systems (with halo mass $\sim10^{14}~\mathrm{M}_\odot$) associated with luminous red galaxies \citep{hand11}.
Others observed systems with halo mass $\sim10^{13}~\mathrm{M}_\odot$ that have characteristic total energies of $7\times10^{61}$~ergs in the circumgalactic medium \citep{plan13,grec15,spac16,spac17,spac18,tani19,pand19}. 
These latter studies aim to detect fossil energy from past AGN activity -- coincidentally of the same order as the cluster feedback energy scale quoted above.
Detection of this fossil energy hinges on the long timescale for radiative cooling in the dark matter halos, which is on the order of a Hubble time for a post shocked gas with entropy in excess of 100 keV (\citealt{voit01, scan04}).
This fossil feedback energy is still a modest fraction of the total energy (and thus SZ) budget, and these studies have yet to find compelling evidence of the AGN contribution.
\citet{gral14} arrived at a similar conclusion when considering massive ellipticals with ongoing radio AGN activity. 
A potentially more fruitful way to find evidence of fossil AGN activity using the SZ effect is through the radial redistribution of the CGM \citep{lebr15}, though arcminute or better resolution is required \citep{tani19}.

Compared to the massive ellipticals discussed above, quasar host galaxies are, on average, less massive ($1-5\times10^{12}~\mathrm{M}_\odot$), corresponding to total CGM energies of $10^{59}-10^{60}$~ergs \citep{sher12,whit12,shen13,dipo14,dipo15,wang15,Efte19, geac19}. At the same time, the feedback energy scale is the same as in the more massive systems ($10^{60}$~ergs), making its detection above the in situ CGM thermal energy more tractable. 
The challenge to measuring the SZ effect in quasar hosts is that the millimeter-wave emission is more complicated in the active systems than in the massive, quiescent ellipticals. 
First attempts at estimating the SZ in quasar hosts include \cite{chat10} and \cite{ruan15}. 
\cite{verd16} used a matched filter approach on \textit{Planck} data to estimate an energy of $\sim4\times10^{61}$~ergs. 
This thermal energy is above expectations of feedback models as it corresponds to $\sim$60 per cent of the bolometric luminosity output of quasars with $L_{\rm bol}\sim1\times10^{46}$~ergs over $10^8$ years.
Furthermore, it is greater than the expected energy from virialization of the highest estimates of dark matter halo mass \citep{rich12}, and as discussed further in this paper, it can likely be attributed to contributions to the SZ signal from correlated neighboring dark matter halos.

\begin{figure}
\includegraphics[width=3.25in]{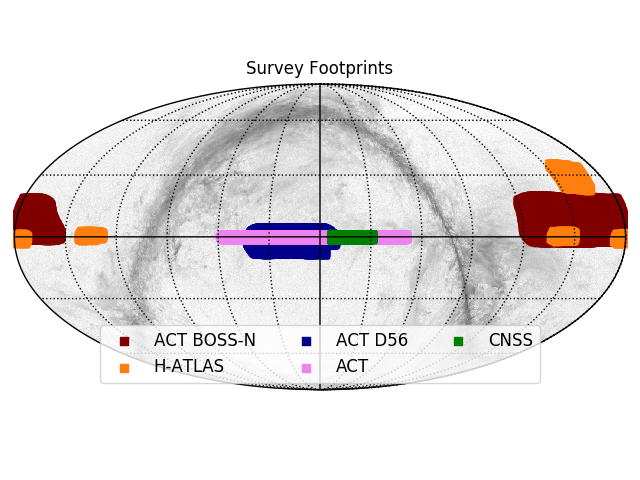}
\caption{Survey footprints for the ACT equatorial, ACTPol D56, ACTPol BOSS-N, H-ATLAS, and CNSS regions in equatorial coordinates. Table~\ref{Nqso} lists the numbers of quasars in each survey field as a function of redshift bin.  \label{qsocoords}}
\end{figure}

\begin{figure}
\includegraphics[width=3.5in]{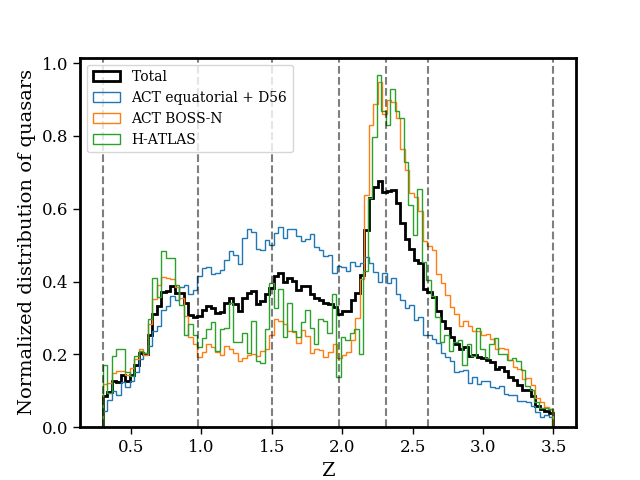}
\caption{Redshift distributions, normalized to form a probability density, of the sum total 109,829 quasars that lie within the ACT equatorial, ACTPol D56, ACTPol BOSS-N, and H-ATLAS footprints. The number of quasars in each survey field is listed as a function of redshift bin in Table~\ref{Nqso}. The number of quasars that lie within ACT equatorial+ACTPol D56 and ACTPol BOSS-N are comparable, while the total number that lie within the H-ATLAS fields is $\sim$10 per cent that of ACT equatorial+ACTPol D56. The shape of the redshift distribution is determined by the SDSS target selection. Vertical dashed lines mark the boundaries of the redshift bins. \label{dndzq}}
\end{figure}

\cite{soer17} fit stacked quasar spectral energy distributions (SED) derived from \textit{Planck} data and found $5\times10^{60}$~ergs at $3-4\sigma$ significance with strong caveats about dependency on dust emission models. 
In \cite{cric16}, we used Atacama Cosmology Telescope (ACT) and \textit{Herschel} data to obtain similar integrated SZ levels to \cite{soer17} with a dependence on the assumed dust emission model. 
\cite{soer17} emphasized the difference between the dust models derived in \cite{cric16} and the \textit{Planck} results; however the large ($5'-10'$) beam of \textit{Planck} means that \cite{soer17} and \cite{verd16} primarily probe dust in the correlated population of dusty star forming galaxies (i.e., clustered sources contributing to the cosmic infrared background; \citealt{hall18}).
The ACT and \textit{Herschel} data at higher--resolution are less affected by the clustered signal on scales beyond the dark matter halo of the host.
They are instead dominated by warmer dust associated with the quasar host itself and the clustered galaxies within the virial radius of the host dark matter halo. 
The resolution impacts the interpretation of the SZ effect as well: for halo masses in the range $10^{12}-10^{13}~\mathrm{M}_\odot$, the \textit{Planck} data are dominated by the 2-halo term associated with correlated large scale structure, whereas the higher resolution ACT and \textit{Herschel} data more directly probe the gas associated with the quasar system and the 1-halo term of correlated structure (\citealt{hill18, hall18, cen15}). We revisit this in the discussion of the implications of our result.

A complementary approach is high--resolution, high--sensitivity observations of the SZ effect associated with individual AGN. 
Recently, such results on both radio and quasar feedback have been published using interferometric data \citep{abdu19,lacy19}.
In the case of quasar feedback, \citet{lacy19} measure an SZ decrement due to quasar winds at the $\sim$3$\sigma$ level, with wind luminosity that is $\sim0.01$ per cent of the bolometric luminosity of the observed hyperluminous quasar at $z\sim1.6$. 

In this work, we present composite quasar SEDs at $z=0.3-3.5$ to study the thermal SZ effect associated with quasar feedback. 
In Section 2 we present the datasets used for the analysis. In Section 3 we describe the measured stacked SEDs of quasars as well as the modelling of the stacked SEDs. Section 4 explores the dependencies of the model, and in Section 5 we further discuss our derived parameters. We summarize and conclude in Section 6. 
For cosmological parameter calculations, such as the comoving distance at a given redshift $z$, we use the default \textit{Planck} 2015 cosmology \citep{plan15} with $H_0 = 67.81$ km s$^{-1}$ Mpc$^{-1}$, $\Omega_m = 0.308$, and $\Omega_{\Lambda} = 0.692$. Throughout the paper, we assess statistical significance with $\chi^2$, number of degrees of freedom, and the probability to exceed (PTE) the $\chi^2$.

\section{Data}
\label{data}

For this analysis, we stack maps from the Atacama Cosmology Telescope, the \textit{Herschel} Space Observatory, and the Very Large Array on the locations of optically-selected quasars from the Sloan Digital Sky Survey (SDSS).
This section first outlines the selection of the quasar samples overlapping the various survey regions, which are largely disjoint. We search for a dependence of the stacked quasar signal on differences in quasar selection between survey regions. We find no such dependence, and detail this in the following section. 
We then describe the maps and survey regions used in the stacking analysis in the subsequent subsections. Table ~\ref{mapstats} gives each map's frequency, beam FWHM, and rms noise values.

\begin{table*}
\centering
\caption{Survey parameters.  \label{mapstats}}

\begin{tabular}{c | c | c | ccc | cc | cc | ccccc}
\hline

Survey & FIRST & CNSS & \multicolumn{3}{c|}{ACT~equatorial}&\multicolumn{2}{c|}{ACTPol~D56} & \multicolumn{2}{c|}{ACTPol~BOSS-N} & \multicolumn{5}{c}{H-ATLAS}\\
\hline
Frequencies (GHz) & 1.4 & 3 & 148 & 220 & 277 & 95 & 148 & 95 & 148 & 600 & 857 & 1200 & 1875 & 3000 \\
Beam FWHMs & 5$^{\prime \prime}$ & 3$^{\prime \prime}$ & 1.4$^\prime$ & 1.0$^\prime$ & 0.9$^\prime$ & 2.1$^\prime$ & 1.4$^\prime$ & 2.1$^\prime$ & 1.4$^\prime$ & 18.1$^{\prime \prime}$ & 25.2$^{\prime \prime}$ & 36.3$^{\prime \prime}$ & 11.4$^{\prime \prime}$ & 13.7$^{\prime \prime}$  \\
RMS noise values & 0.15 & 0.035 & 2.2 & 3.3 & 6.5 & 1.7 & 2.2 & 1.9 & 2.8 & 13.0 & 12.9 & 14.8 & 49 & 44  \\
(mJy/beam) & & & & & & & & & & & & & & \\

\hline
\end{tabular}
\end{table*}

\subsection{Sloan Digital Sky Survey Data Release 14 Quasar Catalog}
\label{qsocatalog}

The quasar sample used in this study is a subset from the SDSS Data Release 14 (DR14, \citealt{pari18}) spectroscopic quasar catalog. 
The DR14 quasar catalog contains all of the quasars observed as a part of SDSS-I/II/III data releases 9, 12, and 14 (DR9, \citealt{pari12}; DR12, \citealt{pari14}; DR14, \citealt{pari17}), and the new quasars that were observed as a part of the SDSS-IV/extended Baryon Oscillation Spectroscopic Survey (eBOSS). 
All objects in the catalog are spectroscopically confirmed as quasars.

We first restrict our quasar sample to the redshift range $0.3<z<3.5$ as this range allows us to avoid the low population tails of the distribution.
Then we apply a cut on the radio emission of the quasars in order to ensure that our objects are radio-quiet by following the methods of \citet{xu99}. 
This study classifies radio loud objects according to the bimodal distribution of the radio 5~GHz luminosity and the [OIII] 5007 $\angstrom$ line luminosity $L_\mathrm{[OIII]}$.
We compute $L_\mathrm{5 GHz}$ from $L_\mathrm{1.4 GHz}$ assuming a synchrotron spectrum with spectral index $-1$, and we compute $L_\mathrm{[OIII]}$ from M$_{2500}$ using the mean relation given in \citet{reye08}. 
M$_{2500}$ is the absolute magnitude at rest-frame 2500 $\angstrom$ for which we use the SDSS K-corrected \textit{i}-band absolute magnitude $M_I(z=2)$ as a proxy \citep{rich06}.
The purpose of ensuring that there are no radio-loud quasars in our sample is twofold. 
Firstly, the majority of the quasar population (90 per cent) is radio quiet, and we are using the SZ effect to probe the potential of the radiative feedback mode for influencing galaxy evolution. 
Secondly, we wish to minimize the contamination of synchrotron emission that is strong in radio-loud objects. 
We apply one final cut to the quasar selection that removes all quasars within 2.5$'$ of bright sources that are detected at $\geq5\sigma$ in the ACT data.

Our final sample includes 109,829 quasars. 
The breakdown of the number of quasars in each field with extractable flux densities is summarized in Table \ref{Nqso}. 
The quasar samples drawn from regions overlapping the combined ACT equatorial and ACTPol D56, the ACTPol BOSS-N, and H-ATLAS are largely independent from one another, with the exception of 1,874 sources that lie within BOSS-N and two of the H-ATLAS GAMA fields, GAMA-12 and GAMA-15.  
The quasar samples constituting the FIRST and CNSS samples contain objects that lie within the other fields. 
The ACT equatorial and ACTPol D56 survey fields partially overlap. 
Figure \ref{qsocoords} displays the three ACT regions,
as well as the H-ATLAS and CNSS regions.

Figure~\ref{dndzq} shows the normalized redshift distribution of the combined quasar samples overlapping all of the ACT regions and the H-ATLAS regions, with dashed vertical lines at the redshift boundaries of each bin. 
Shown separately are the redshift distributions of the quasars lying within each of the survey fields.
Each curve is normalized to form a probability distribution. 
The strong redshift dependence of the quasar distribution reflects the selection function and the targeting priorities of the SDSS survey (\citealt{myer15, pari18}). 
The quasar samples overlapping the ACT equatorial region and ACTPol D56 overlap with the SDSS region Stripe 82 (S82) that covers the Celestial Equator, though D56 extends beyond S82 in declination. 
The H-ATLAS and ACTPol BOSS-N regions are above the Galactic plane (see Figure \ref{qsocoords}). 
The quasar selection difference in these other survey fields is evident in the respective quasar redshift distributions in Figure~\ref{dndzq}.
The fields above the Galactic plane (orange and green histograms in Firgure~\ref{dndzq}) have the most quasar coverage from SDSS ~III, which focused on quasars at $z\geq2.15$ in order to access the Lyman-$\alpha$ forest \citep{pari18}.

In our construction of the radio through far-infrared quasar SEDs we tested for any discrepancies in stacked flux density at a given frequency that may be caused by differences the quasar samples in each of the survey footprints. 
We matched the quasar samples from the different fields in their absolute magnitude distributions and in their $r-i$ and $i-z$ color distributions and re-calculated their broadband SEDs. 
We find no difference in flux density within the uncertainties of the data in these SEDs compared to those made from the full, combined quasar samples from each field.
When a frequency band has data spanning multiple fields, as with ACTPol D56 and BOSS-N at 95~GHz and 148~GHz, the flux density estimates also agree between fields (however, see caveat regarding \textit{Herschel} SPIRE data in Section~\ref{hershdata}). 
We also find consistency between flux densities at 1.4~GHz derived from stacking all of the quasars in our sample and those derived from using quasars in the individual fields, including when limiting to only the quasars that lie within the CNSS Pilot survey.
We therefore continue the analysis with the full quasar sample with the numbers in each field as reported in Table~\ref{Nqso}.

\begin{table*}
\centering
\caption{Number of Quasars in each redshift bin and survey field. The numbers of quasars listed for ACT equatorial/ACTpol D56, ACTPol BOSS-N, and H-ATLAS fields are all independent, with the exception of 1,874 sources overlapping BOSS-N and GAMA-12/GAMA-15, while the quasars in FIRST and CNSS overlap with the other fields. The numbers in parentheses in the ACT equatorial column are those that also lie within ACTPol D56. The numbers in parentheses in the H-ATLAS column are those that overlap with ACTPol BOSS-N. The total represents the total number of independent objects in each redshift bin. \label{Nqso}}

\begin{tabular}{lcccc | cc | c}
\hline

 & ACT equatorial & ACTPol~D56 & ACTPol BOSS-N & H-ATLAS & FIRST & CNSS & Total \\
$z$ range & (148, 218, 277~GHz) & (95, 148~GHz) & (95, 148~GHz) & (600, 857, 1200, 1875, 3000~GHz) & (1.4~GHz) & (3~GHz) &  \\ 
 \hline
 0.3-0.98 & 3439 (2222) & 7518 & 8678 & 1197 (393) & 17162 & 579 & 18217 \\
 0.98-1.47 & 4695 (3380) & 11454 & 5068 & 784 (247) & 18277 & 478 & 18374 \\
 1.47-1.91 & 4671 (3408) & 11164 & 5360 & 833 (231) & 18307 & 471 & 18389 \\
 1.91-2.2 & 3289 (2322) & 8239 & 8383 & 989 (294) & 18222 & 401 & 18284 \\
 2.2-2.59 & 2171 (1368) & 5069 & 11312 & 1440 (374) & 18213 & 415 & 18250 \\
 2.59-3.5 & 2659 (1742) & 5328 & 11233 & 1172 (335) & 18274 & 495 & 18315 \\
 
\hline
\end{tabular}

\end{table*}

\subsection{Atacama Cosmology Telescope (ACT)}
\label{actdata}

The Atacama Cosmology Telescope (ACT) is a 6-m telescope located in the Atacama desert in northern Chile \citep{fowl07}. 
Its first of three cameras was the Millimeter Bolometric Array Camera (MBAC) that observed at 148, 218, and 277 GHz, making it well equipped for mapping the cosmic microwave background (CMB) and for observing the SZ effect \citep{swet11}. 
This camera took observations from 2007-2010, and with it the telescope mapped over 1000 deg$^2$ of the sky. 
Details concerning the ACT data reduction, map calibrations, and descriptions of the beams are given in \citet{dunn13}, \citet{haji11}, and \citet{hass13}. 
The second-generation ACT survey was conducted with the polarization-sensitive ACTPol receiver. 
The ACTPol survey took place from 2013-2015 in frequency bands 95~GHz and 148~GHz (\citealt{niem10, thor16}). 
\cite{loui17} give a description of the ACTPol data reduction, calibration, and beams.
The third generation of ACT has been ongoing since 2016 via the Advanced ACTPol experiment (AdvACT; \citealt{hend16}). 

In this study, we use the equatorial maps in all three frequency bands from the ACT/MBAC survey, which cover 340~deg$^2$ ($-55^\circ$ < RA < $58^\circ$; $-1.5^\circ$ < dec < $1.5^\circ$). The 148~GHz and 218~GHz data were taken from 2009 to 2010. The 277~GHz data are only from 2010. From ACTPol we use the equatorial maps in the region known as D56 and the maps in the region called BOSS-N. These fields were observed in the second and third seasons of ACTPol operation. The D56 region covers $\sim$700~deg$^2$ (approximately ${-10^\circ < \mathrm{RA} < 40^\circ}$; ${-9^\circ < \mathrm{dec} < 5^\circ}$), and BOSS-N covers over $\sim$2000~deg$^2$ (approximately $130^\circ < \mathrm{RA} < 240^\circ$; $-5^\circ < \mathrm{dec} < 20^\circ$). 
The ACTPol survey strategy is described by \citet{debe16}.  Finally, no data from the ongoing Advanced ACTPol survey are used in this study. All of the ACT maps have been matched-filtered and put into units of Jy beam$^{-1}$ such that the flux density recovered from each pixel is representative of that of a point source in that position (\citealt{meli06, marr11}). 

The absolute calibration of the ACT equatorial 148~GHz and 218~GHz bands was determined via a cross-correlation analysis with \textit{WMAP} \citep{haji11}, while the 277~GHz maps were calibrated using Uranus and Saturn. 
The beam full widths at half maximum (FWHM) for the ACT 148~GHz, 218~GHz, and 277~GHz data are 1.34$'$, 1.01$'$, and 0.89$'$, respectively.
Additionally, all of the data have uncertainties associated with the beam measurements and map making procedures.
The resulting calibration uncertainties in the 148~GHz and 218~GHz maps are 3 per cent and 5 per cent, respectively, and these two bands are correlated at a level of 3 per cent due to their calibration analysis procedure \citep{das14}. 
The 277~GHz flux densities have a calibration uncertainty of 15 per cent, and it is not correlated with the other bands. (See \citealt{gral14} for details.)

The ACTPol experiment was equipped with three detector arrays. The first two observed at 148~GHz, while the third observed simultaneously at 95 and 148~GHz. 
Observations of Uranus are used to characterize the beams.
The 148~GHz beam sizes at FWHM are 1.37$'$, 1.33$'$, and 1.58$'$ for detector arrays 1, 2, and 3, respectively.
The 95~GHz beam of the third detector array has a FWHM of 2.13'.
All of the calibration uncertainties are on the order of 2 per cent as determined by cross correlation with \textit{Planck}. 
The ACTPol calibrations have not yet been finalized and may change at the $\sim$1 per cent level, but our results in this paper are robust against these changes.
More details of the ACTPol data and instrument can be found in \citet{thor16} and \citet{loui17}.

\subsection{\textit{Herschel} Space Observatory}
\label{hershdata}

To measure the thermal dust emission spectrum of the stacked quasar SEDs, we use far-infrared data from the \textit{Herschel} Space Observatory. 
The \textit{Herschel} mission's far-infrared observations span 2009 to 2013. 
It was equipped with three instruments: the Photodetector Array Camera and Spectrometer (PACS, \citealt{pogl10}), the Spectral and Photometric Imaging REceiver (SPIRE, \citealt{grif10}), and the Heterodyne Instrument for the Far Infrared (HIFI, \mbox{\citealt{graa10}}). All three instruments had spectrometers, and PACS and SPIRE were equipped with cameras. 

We use images from the \textit{Herschel}-Astrophysical Terahertz Large Area Survey (H-ATLAS; \citealt{eale10}) that was carried out using the PACS and SPIRE instruments. The H-ATLAS survey mapped over 600 deg$^2$ of the sky in five photometric bands: 100~$\mu$m and 160~$\mu$m using the PACS instrument, and 250~$\mu$m, 350~$\mu$m, and 500~$\mu$m using the SPIRE instrument. 
Specifically, we use the $\sim$180~deg$^2$ patch of the survey that is centered on the North Galactic Pole (NGP; \citealt{smit17}) and the three 60~deg$^2$ equatorial regions that coincide with the Galaxy And Mass Assembly (GAMA) survey \citep{vali16}. 
These three fields are centered on right ascensions 9h, 12h, and 15h and each extends 4 degrees in declination. 
All of the maps we use came from the H-ATLAS team as a part of data release two of the survey. 
Details of the maps processing and properties can be found in \citet{vali16}. 
All of the maps have been background-subtracted using the \textit{Nebuliser}  algorithm to remove any large-scale ($>3'$) Galactic or extragalactic emission. 

We do not use the available SPIRE data from the \textit{Herschel} Multi-tiered Extragalactic Survey (HerMES) or \textit{Herschel} Stripe 82 Survey (HerS) due to discrepancies in the flux densities of stacked quasars in the HerMES/HerS vs. the H-ATLAS fields. 
The most relevant HerMES field is the HerMES Large Mode Survey (HeLMS), which covers 270 deg$^2$ and overlaps Stripe 82.
Specifically, we find that the SPIRE points from the HeLMS/HerS fields tend to be lower in flux density, particularly in the 250~$\mu$m band where the HeLMS/HerS points are up to 50 per cent lower in flux than the H-ATLAS points. As a result, the HeLMS/HerS SPIRE data yield different spectral shapes with cooler dust temperatures than those indicated by the combined SPIRE and PACS data from the H-ATLAS fields.
We tested for any difference in quasar target selection as described in Section~\ref{qsocatalog} that may be driving these discrepancies, and found none.
The H-ATLAS maps we use have been matched-filtered, while the HeLMS and HerS maps have not. For the latter, we perform our own background subtraction in order to account for correlation emission from dusty galaxies. It is possible that the HeLMS and HerS maps still contain a larger contamination from clustered sources, which is one reason for continuing with only the H-ATLAS datasets.
For the purposes of constraining the dust spectrum in this study, we continue the analysis using the H-ATLAS SPIRE and PACS data only. 
For completeness we performed the analysis using the HeLMS and HerS SPIRE data only (no PACS), and we do not find any significant discrepancy ($\lesssim$1$\sigma$) in our marginalized SZ amplitude. 

In order to measure the Wien part of the dust spectrum, we use images from the \textit{Herschel} PACS instrument at 100~$\mu$m and 160~$\mu$m with FWHM beam sizes of 11.4$''$ and 13.7$''$, respectively. 
The PACS maps have units of Jy~pixel$^{-1}$. 
We convert the recovered flux densities to Jy~beam$^{-1}$ by performing aperture photometry with a circular aperture of radius equal to the FWHM of the respective band, and then applying an aperture correction using the encircled energy function \citep{vali16}. 
The primary uncertainty in the flux densities from PACS is derived from calibration uncertainties of stars and asteroids, and the total calibration error is estimated to be 5 per cent.

The FWHM of SPIRE beams are derived from observations of Neptune and are 18.1$''$, 25.2$''$, and 36.6$''$ for the 250~$\mu$m, 350~$\mu$m, and 500~$\mu$m bands, respectively \citep{grif13}. The flux sensitivity of the \textit{Herschel}-SPIRE bands is 13.0, 12.9, and 14.8~mJy~beam$^{-1}$ for the 250~$\mu$m, 350~$\mu$m, and 500~$\mu$m bands, respectively, with the confusion noise contributing 8 mJy~beam$^{-1}$ at all wavelengths. 
The SPIRE maps we use have been matched-filtered using the SPIRE point spread function and are in units of Jy~beam$^{-1}$. 
The flux density calibration uncertainties derived from the Neptune observations total 7 per cent. 
The calibration uncertainty of all three bands is correlated at the level of 5 per cent.
As a check on our SPIRE and PACS flux density determination, we tested that we accurately recover the fluxes of detected sources from available H-ATLAS catalogs, and find that our flux calculations agree with the catalog values within the catalog uncertainties. 

\subsection{The VLA FIRST Survey}

The Faint Images of the Radio Sky at Twenty-one centimeters (FIRST) survey from the Very Large Array (VLA) mapped over 10,000 deg$^2$ of the sky at 1.4 GHz (21-cm) with a resolution of $5''$, typical rms of 0.15 mJy, and a detection threshold of 1 mJy \citep{beck95}. 
We use these data to obtain the stacked flux density of our quasars at 1.4 GHz. 
FIRST overlaps with all of our survey fields, and we obtain $5'\times5'$ image thumbnail cutouts centered on 108,455 quasars.
(There are quasars in the sample for which a valid FIRST flux density could not be recovered.) The data are calibrated in units of Jy~beam$^{-1}$, and the pixel sizes are $1.8''$. We extract the flux density from the pixel corresponding to the SDSS (RA, dec) coordinate of each quasar.

\subsection{The Caltech-NRAO Stripe 82 Survey}

The Caltech-NRAO Stripe 82 Survey (CNSS) has made public its pilot survey that mapped 50 deg$^2$ of the sky at 3 GHz with a resolution of $3''$ using the Jansky VLA \citep{mool16}. 
The combined map has a median rms noise of $\sim$35~$\mu$Jy.
Using this pilot survey in conjunction with FIRST, we can begin to constrain the shape of the radio spectrum of our sample. 
We obtain $1'\times1'$ thumbnail cutouts of 2,838 quasars in the CNSS field. 
The cutouts are in units of Jy~beam$^{-1}$, and the pixel sizes are $0.75''$. We extract the flux density from the pixel corresponding to the SDSS (RA, dec) coordinate of each quasar.

\begin{figure*}
\includegraphics[width=7in]{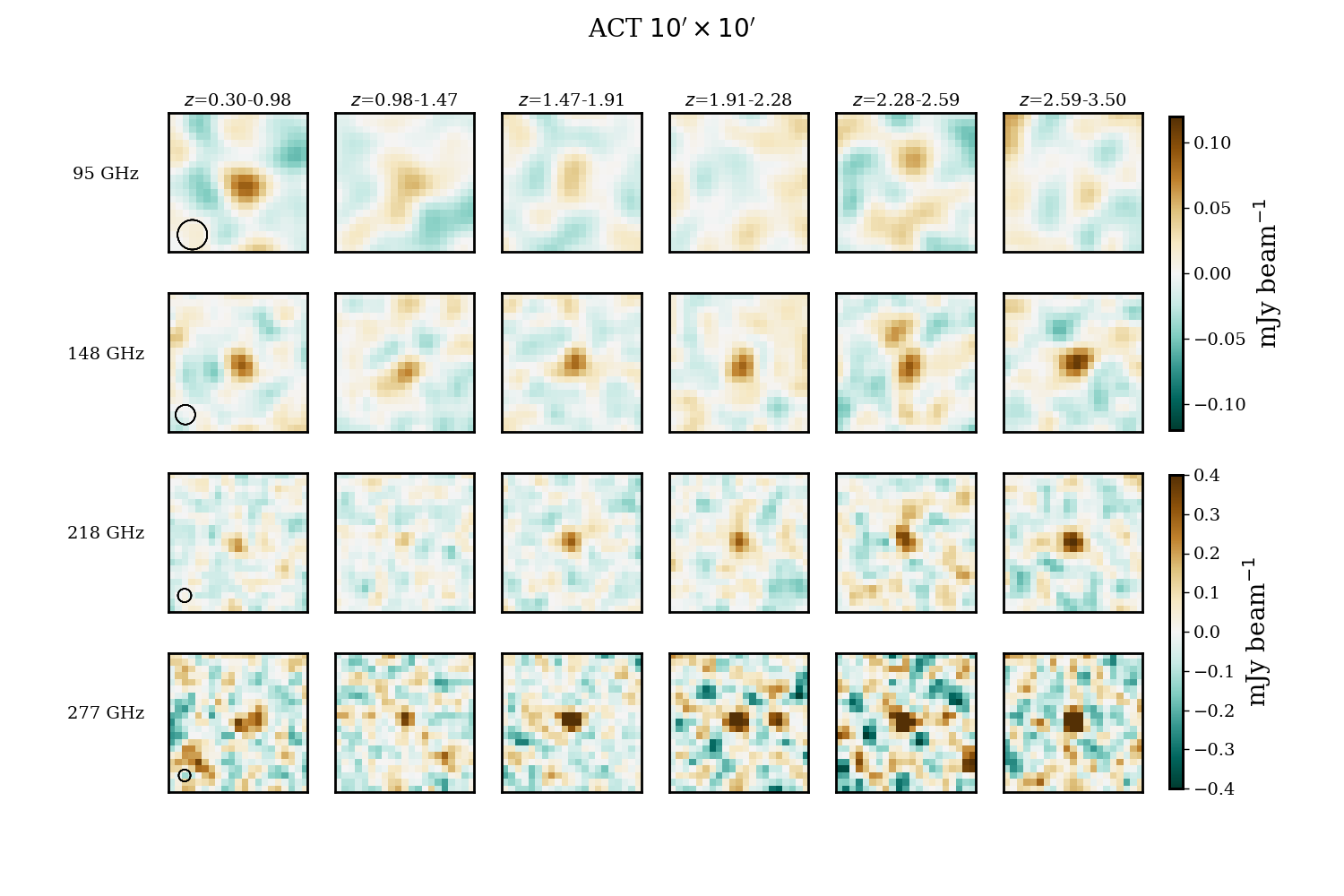}
\caption{ACT $10'\times10'$ stacked thumbnails as a function of frequency and redshift. Black circles in the lower left corner of the cutouts in the first column indicate the FWHM of each band's beam. The flux scale of the 95~GHz and 148~GHz thumbnails is indicated by the top colorbar, while the 218~GHz and 277~GHz thumbnails' flux scale is indicated by the bottom colorbar. \label{ACTthumbs}}
\end{figure*}

\begin{figure*} 
\includegraphics[width=7in]{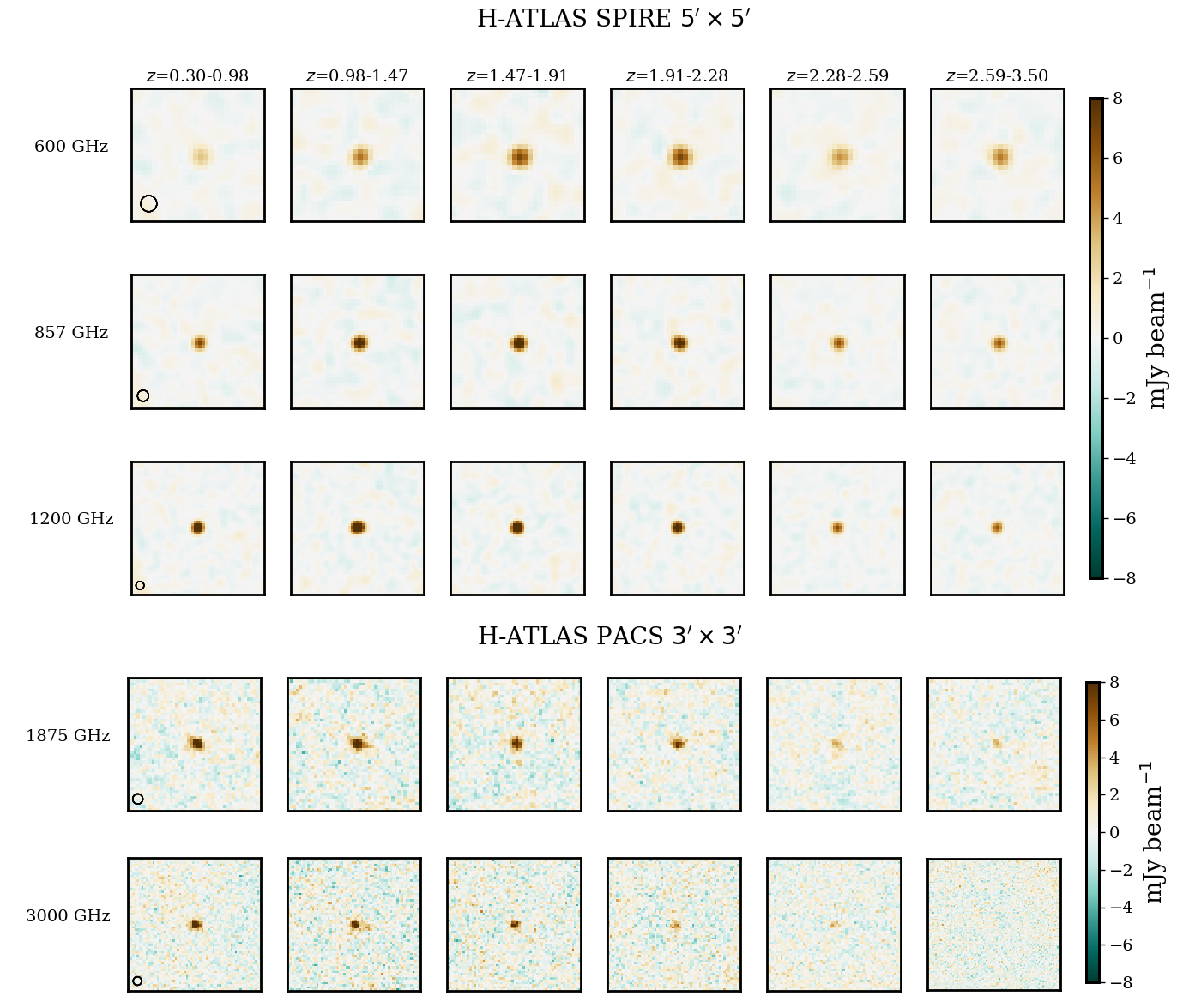}
\caption{H-ATLAS SPIRE $5'\times5'$ and PACS $3'\times3'$ stacked thumbnails as a function of frequency and redshift. The PACS thumbnails are converted from mJy~pixel$^{-1}$ to mJy~beam$^{-1}$ by multiplying by the beam area in square arcseconds divided by the square of the pixel sizes. Black circles in the lower left corner of the cutouts in the first column indicate the FWHM of each band's beam. The SPIRE thumbnails are displayed on a common flux scale indicated by the top color bar, and similarly the flux scale of the PACS thumbnails is given by the bottom colorbar. \label{HATLASthumbs}}
\end{figure*}

\begin{table*}
\centering
\begin{threeparttable}
\caption{Measured flux densities for the quasars in each frequency band and redshift bin as described in Section~\ref{fdmeasurements}.\label{fd}}

\begin{tabular}{lcccccc}
\hline
                                    & $z$ = 0.30 - 0.98 & $z$ = 0.98 - 1.47 & $z$ = 1.47 - 1.91 & $z$ = 1.91 - 2.28 & $z$ = 2.28 - 2.59 & $z$ = 2.59 - 3.50 \\
 \hline
 S$_\mathrm{1.4}^\mathrm{a}$ (mJy)  & 0.05 $\pm$ 0.02   & 0.04 $\pm$ 0.004   & 0.031 $\pm$ 0.002 & 0.024 $\pm$ 0.002 & 0.020 $\pm$ 0.001 & 0.021 $\pm$ 0.001 \\
 S$_\mathrm{3}^\mathrm{a}$ (mJy)    & 0.03 $\pm$ 0.03   & 0.014 $\pm$ 0.008 & 0.015 $\pm$ 0.005 & 0.012 $\pm$ 0.005 & 0.009 $\pm$ 0.008 & 0.016 $\pm$ 0.005 \\
 S$_\mathrm{95}$ (mJy)              & 0.08 $\pm$ 0.02   & 0.05 $\pm$ 0.02   & 0.04 $\pm$ 0.02   & -0.02 $\pm$ 0.02  & 0.004 $\pm$ 0.020 & 0.008 $\pm$ 0.019 \\
 S$_\mathrm{148}^\mathrm{b}$ (mJy)  & 0.08 $\pm$ 0.02   & 0.04 $\pm$ 0.02   & 0.08 $\pm$ 0.02   & 0.07 $\pm$ 0.02   & 0.08 $\pm$ 0.02   & 0.11 $\pm$ 0.02   \\
 S$_\mathrm{218}^\mathrm{b}$ (mJy)  & 0.22 $\pm$ 0.05   & 0.17 $\pm$ 0.04   & 0.33 $\pm$ 0.06   & 0.34 $\pm$ 0.05   & 0.37 $\pm$ 0.07   & 0.52 $\pm$ 0.06   \\
 S$_\mathrm{277}^\mathrm{c}$ (mJy)  & 0.41 $\pm$ 0.12   & 0.39 $\pm$ 0.10   & 0.80 $\pm$ 0.15   & 0.85 $\pm$ 0.16   & 0.85 $\pm$ 0.19   & 1.29 $\pm$ 0.23   \\
 S$_\mathrm{600}^\mathrm{d}$ (mJy)  & 2.5 $\pm$ 0.3     & 4.2 $\pm$ 0.5     & 5.5 $\pm$ 0.5     & 5.6 $\pm$ 0.5     & 3.5 $\pm$ 0.4     & 4.1 $\pm$ 0.4     \\
 S$_\mathrm{857}^\mathrm{d}$ (mJy)  & 5.5 $\pm$ 0.5     & 8.0 $\pm$ 0.7     & 8.7 $\pm$ 0.8     & 7.7 $\pm$ 0.7     & 5.2 $\pm$ 0.5     & 5.0 $\pm$ 0.5     \\
 S$_\mathrm{1200}^\mathrm{d}$ (mJy) & 11.6 $\pm$ 1.0    & 13.7 $\pm$ 1.2    & 13.0 $\pm$ 1.1    & 10.5 $\pm$ 0.9    & 6.4 $\pm$ 0.5     & 6.0 $\pm$ 0.5     \\
 S$_\mathrm{1875}^\mathrm{e}$ (mJy) & 17.6 $\pm$ 1.9    & 15.6 $\pm$ 1.8    & 13.5 $\pm$ 1.6    & 10.2 $\pm$ 1.6    & 6.1 $\pm$ 1.2     & 3.4 $\pm$ 1.4     \\
 S$_\mathrm{3000}^\mathrm{e}$ (mJy) & 14.8 $\pm$ 1.4    & 12.2 $\pm$ 1.5    & 10.0 $\pm$ 1.4    & 6.1 $\pm$ 1.3     & 3.6 $\pm$ 1.2     & 1.9 $\pm$ 1.2     \\
\hline
\end{tabular}

\begin{tablenotes}
\small
\item $^\mathrm{a}$ Reported flux densities are median values and the uncertainties are the standard errors on the medians.
\item $^\mathrm{b}$ Calibration uncertainties of 3 per cent(5 per cent) are added to the 148(218)~GHz flux densities, as well as a 3 per cent correlated component.
\item $^\mathrm{c}$ A calibration uncertainty of 15 per cent is added in quadrature.
\item $^\mathrm{d}$ A calibration uncertainty of 7 per cent is added in quadrature, as well as a 5 per cent correlated component.
\item $^\mathrm{e}$ A calibration uncertainty of 5 per cent is added in quadrature.
\end{tablenotes}

\end{threeparttable}
\end{table*}

\section{Stacked Spectral Energy Distributions}

\subsection{Stacking to recover the average flux densities of quasars}
\label{fdmeasurements}

The quasars in our sample are optically bright, but have far-infrared, millimeter, and radio flux densities that fall below the detection limits of the surveys used in this work. 
We perform a stacking analysis to measure the mean flux density of the quasars in each redshift bin at ACT and H-ATLAS map frequencies. 
Our stacking algorithm entails cutting out square patches (thumbnails) of the maps around the (RA, dec) coordinate of each quasar in each redshift bin and then computing an inverse variance-weighted average of all thumbnails in a bin. 
Stacking is equivalent to taking the cross-correlation of an intensity map with a catalog (\citealt{mars09}; \citealt{vier13}). 
Stacking a confusion-dominated or Gaussian noise-dominated map, like the ACT and \textit{Herschel} maps, on the locations of sources allows us to obtain an unbiased maximum likelihood estimate of the average flux of the sources in the catalog. 

We stack the ACT and \textit{Herschel} maps on the locations of quasars in six redshift bins, divided such that there is approximately the same number of quasars in each bin, $N\sim 18300$, as defined by the concatenation of quasars overlapping all of the survey fields used in this study.
We mean-subtract each of the complete maps before stacking.  For each redshift bin, the stacked signal is constructed by taking the inverse variance weighted average of the thumbnails centered on all quasars in that redshift bin:
\begin{equation}
\bar{S_{\nu}^{\delta}} = \frac{\sum\limits_{i}^{N_{\delta}}w_{i,\nu}s_{i,\nu}}{\sum\limits_{i}^{N_{\delta}}w_{i,\nu}}.
\label{eq:stackeq}
\end{equation}
Equation~\ref{eq:stackeq} gives the mean stacked intensity map $\bar{S_{\nu}^{\delta}}$ of the redshift bin $\delta$ containing $N_{\delta}$ sources, each contributing individual map intensity $s_{i,\nu}$ in the frequency band $\nu$. 
For each quasar thumbnail, the inverse variance weights $w_{i,\nu}$ are determined on a pixel-by-pixel basis from the hit (counts) maps for the ACT and PACS data and from the square of the error maps for the SPIRE data. 
The ACT flux densities are also multiplied by correction factors to account for the probability of each source falling anywhere within the $0.5'$-square pixels.
These factors are 1.02, 1.06, 1.10, and 1.14 for the 95, 148, 218, and 277 GHz bands, respectively \citep{gral14}. 
Stacked $10'\times10'$ thumbnails for the ACT frequencies are given in Figure~\ref{ACTthumbs}.
Stacked $5'\times5'$ thumbnails for the H-ATLAS SPIRE frequencies and stacked $3'\times3'$ thumbnails for the H-ATLAS PACS frequencies are given in Figure~\ref{HATLASthumbs}. 

For radio data from FIRST and CNSS, we follow a modified stacking procedure. Firstly, the data are combined after extracting  flux densities from the thumbnail cutouts of individual quasars (instead of combining the thumbnails and extracting the stacked flux density from the aggregate thumbnail). Secondly, instead of the weighted mean, we compute the median flux density as the representative value for the population in each redshift bin. 
The reason for this is the presence of significant positive outliers in the radio data. (For instance, the brightest source in the lowest redshift bin is 1600 times brighter at 1.4~GHz than the median source.) 
These outliers bias the weighted mean to high values, up to six times the median, and significantly increase the associated error.
We therefore find that the median is a more robust estimate of the stacked flux density in each redshift bin. 
There is also a positive tail of non-outliers, so if we could hypothetically separate the outliers in a non-arbitrary way, the median estimator would differ from the putative mean. 
To investigate the impact of using the median instead of a weighted mean, we compute the medians of the higher frequency data, finding that the median values scatter within 1$\sigma$ about the means without bias.
We also compute the weighted mean of the 95~GHz data in the lowest redshift bin (where the bright radio tail is most pronounced) excluding the 4 per cent of quasars with $S_{1.4}>1$~mJy (the majority of the positive flux density tail). This lowers the 95~GHz stack by $3$~$\mu$Jy, which is 15 per cent of the statistical error and which produces a negligible impact in conclusions regarding excess millimeter emission and the SZ effect.

The stacked flux densities and uncertainties for each redshift and frequency are given in Table~\ref{fd}. 
The average flux densities for frequency bands with $\nu \geq 95$~GHz are computed using Equation~\ref{eq:stackeq}.
The uncertainties are derived using a bootstrap algorithm that randomly samples the same number of objects in a given redshift bin with replacement.
For each redshift bin and frequency band, we draw 100 bootstrapped samples with the same number of sources as our science SEDs and compute the average flux density using Equation~\ref{eq:stackeq}.
The uncertainty is then the standard deviation of 100 averages computed from the bootstrapped flux densities. 
We also add in quadrature any calibration uncertainties and compute a covariance matrix as referenced by the superscripts in Table~\ref{fd}.
The FIRST 1.4~GHz and CNSS 3~GHz flux densities and errors are computed as the median and standard error on the median (1.235$\sigma_\mathrm{dev}$/$\sqrt{N}$, where $\sigma_\mathrm{dev}$ is the standard deviation) of each object's measured flux density in a given redshift bin.
We find these uncertainties are consistent with bootstrapped errors. 

We test for potential bias in the stacked flux via a "null" stacking analysis: stacking randomly drawn positions distributed equally over the sky coverage of each of the survey fields. 
We draw 500 null stacks with equal numbers of random sky positions as there are quasars in the first redshift bin for each band. 
All of the stack procedures remain the same as for stacking on the quasar locations.
In the case of the FIRST and CNSS fields we take the median and the standard error on the median. 
The means of these 500 random stacked fluxes for each band are scattered around zero in an unbiased manner, and all except the 277~GHz flux are found to be consistent with zero within their statistical uncertainty. The 277 GHz point is 1.5$\sigma$ below zero. 
Testing against zero flux density in all 11 bands (11 degrees of freedom) yields a $\chi^2$ = 5.6 with PTE = 0.89.

By using matched-filtered maps, we mitigate bias on scales larger than the beam that would be due to background emission from sources in the map that may be correlated with the quasars. 
Such signals have been seen in cross-correlations of quasar catalogs with \textit{Herschel}-SPIRE data (\citealt{wang15, hall18}). 
The matched-filtering imposes a high-pass filter to remove large scale modes from the data, and therefore bias from background correlation on scales greater than the size of the beam is reduced.  However, as discussed below,  contributions to the measured flux may still arise from correlated structure on beam scales.

\begin{table*}
\centering
\caption{Marginalized parameters derived from the 50th percentile of the posterior distributions from the MCMC analysis. The fit parameters include the radio spectral index $\alpha$, the coupling fraction of the SZ effect $f (\tau_8^{-1} \mathrm{per cent})$, and the warm and cold components of the two temperature modified blackbody dust spectra parameterized by the infrared luminosities $L_{\rm IR}$ and the dust temperatures $T_d$. The total thermal energy $E_{\rm th}$, total infrared luminosity $\log(L_{\rm IR,tot})$, and effective dust temperature $T_{d,eff}$ columns are calculated from the fit parameters. The 1$\sigma$ statistical uncertainties are given by the 16th and 84th percentiles and propagated accordingly for the calculated values. \label{results}}

\begin{tabular}{lccccccccc}
\hline
 $z$ range   & $\alpha$       & $f (\tau_8^{-1} \%)$ & $E_{\rm th}$ (ergs)                     & $\log(L_{\rm IR,c} [L_\odot])$ & $T_{d,c} (K)$  & $\log(L_{\rm IR,w} [L_\odot])$ & $T_{d,w} (K)$  & $\log(L_{\rm IR,tot} [L_\odot])$ & $T_{d,eff}$    \\
 \hline
             &                      &                      &                                     &                            &                &                            &                &                              &                \\
 0.30 - 0.98 &                      &                      & (4.1$^{+0.8}_{-1.1})\times10^{+59}$ & 11.3$^{+0.1}_{-0.1}$       &                & 10.3$^{+0.8}_{-1.2}$       &                & 11.6$^{+0.1}_{-0.04}$         & 37$^{+1}_{-2}$ \\
             &                      &                      &                                     &                            &                &                            &                &                              &                \\
 0.98 - 1.47 &                      &                      & (9.3$^{+3.2}_{-2.4})\times10^{+59}$ & 11.7$^{+0.1}_{-0.1}$       &                & 12.0$^{+0.2}_{-0.2}$       &                & 12.3$^{+0.1}_{-0.1}$         & 60$^{+4}_{-4}$ \\
             &                      &                      &                                     &                            &                &                            &                &                              &                \\
 1.47 - 1.91 & -0.8$^{+0.2}_{-0.2}$ & 4.4$^{+0.9}_{-1.1}$  & (1.7$^{+0.3}_{-0.4})\times10^{+60}$ & 11.9$^{+0.1}_{-0.1}$       & 34$^{+1}_{-1}$ & 12.2$^{+0.2}_{-0.2}$       & 74$^{+6}_{-6}$ & 12.6$^{+0.1}_{-0.1}$         & 61$^{+5}_{-4}$ \\
             &                      &                      &                                     &                            &                &                            &                &                              &                \\
 1.91 - 2.28 &                      &                      & (2.4$^{+0.5}_{-0.6})\times10^{+60}$ & 11.9$^{+0.1}_{-0.1}$       &                & 12.4$^{+0.2}_{-0.2}$       &                & 12.7$^{+0.1}_{-0.1}$         & 63$^{+5}_{-5}$ \\
             &                      &                      &                                     &                            &                &                            &                &                              &                \\
 2.28 - 2.59 &                      &                      & (2.6$^{+0.5}_{-0.7})\times10^{+60}$ & 11.8$^{+0.1}_{-0.1}$       &                & 12.2$^{+0.2}_{-0.2}$       &                & 12.6$^{+0.1}_{-0.1}$         & 62$^{+5}_{-5}$ \\
             &                      &                      &                                     &                            &                &                            &                &                              &                \\
 2.59 - 3.50 &                      &                      & (4.1$^{+0.8}_{-1.0})\times10^{+60}$ & 12.0$^{+0.1}_{-0.1}$       &                & 12.1$^{+0.2}_{-0.3}$       &                & 12.7$^{+0.1}_{-0.1}$         & 57$^{+4}_{-4}$ \\

\hline
\end{tabular}

\end{table*}

\subsection{Modelling the stacked SEDs}

We compute model spectral energy distributions as the addition of a synchrotron emission spectrum, a thermal dust emission spectrum as the sum of two modified blackbodies, additional emission components from CO lines entering the ACT bands, and the spectral distortion from the SZ effect. 

The origin of radio emission in radio quiet quasars remains a topic of some debate (\citealt{laor08, zaka16, hwan18, laor19, pane19}), and quantifying the radio spectral indices of a statistical sample of radio-quiet quasars has not been done. 
In this study, we include the stacked radio flux densities at 1.4 and 3~GHz because if there is significant synchrotron and/or free-free emission in the quasar population, then it has a potential to be present at 95~GHz where we also expect to see a decrement in the emission due to the SZ effect. 
We model the synchrotron component as a power law with respect to frequency, $S_{\nu} \propto \nu^{\alpha}$. 
The power law index $\alpha$ is a negative value in the range $-1.5 < \alpha < -0.5$ consistent with optically thin synchrotron emission. 
We also explore the possibility of flat spectrum ($\alpha > -0.5$) radio emission \citep{dezo10}. 

Galaxies have a distribution of dust temperatures that depends on the sizes and distribution of the dust and on the temperatures of the heating sources. 
The total infrared SED is therefore a sum of many modified blackbodies with different temperatures from all of the dust components. 
Modelling of high-redshift AGN, ultra-luminous IR galaxies, and other local star-forming galaxies has shown that at least two modified blackbodies are needed to describe the full infrared SED (\citealt{dunn01, farr03,kirk12}). 
In our base model, the dust spectrum is thus modeled as the sum of two modified blackbodies. 
Each modified blackbody is computed as a function of rest frame frequency $\nu$ with the functional form
\begin{equation}
    S_{\mathrm{dust}}(\nu, z, L_{\rm IR}, \beta, T_d) = \frac{L_{\rm IR}}{4\pi (1+z)d_M^2(z)}\frac{((1+z)\nu)^{\beta}B_{(1+z)\nu}(T_d)}{\int \nu'^{\beta}B_{\nu'}(T_d)d\nu'}.
\end{equation}
$B(T_d)$ is the Planck blackbody function. The temperature of this optically thin modified blackbody spectrum is defined as $T_d$ and the emissivity given by $\beta$. 
The integral is taken from $8-1000$~$\mu$m in the rest frame in order to normalize the spectrum to the infrared bolometric luminosity $L_{\rm IR}$ of the quasar over the same wavelength range as is standard \citep{kenn98}. 
The comoving distance $d_M(z)$ at the redshift of the quasar is used to generate the SED in terms of the observed frame flux density. 
In our base model, we fit the dust spectrum with the sum of two modified blackbodies. 

The thermal SZ effect \citep{suny70} is a distortion of the spectrum of CMB radiation passing through the hot plasma surrounding the quasar. The SZ effect is proportional to the the volume integrated thermal pressure of the plasma's electron gas, $\int p_e dV$. 
The integrated thermal pressure due to quasar winds is directly proportional to the thermal energy output of the quasar, $E_{\rm th}$, and it is this quantity that we wish to probe as a function of redshift. 
Integrating the flux density over the solid angle of the source, we can write the SZ contribution as: 
\begin{equation}
\label{eq:Ssz}
    S_{\rm SZ}(\nu,z, \int p_edV) = I_0 g(x) \frac{\sigma_T}{m_ec^2}\frac{\int p_edV}{D_A^2(z)},
\end{equation}
where $I_0 = 2(k_BT_{\rm CMB})^3/(hc)^2$ and $g(x)$ is the nonrelativistic frequency dependence of the SZ effect,
\begin{equation}
     g(x \equiv h\nu/k_BT_{\rm CMB}) = \frac{x^4e^x}{(e^x-1)^2}\left( x \frac{e^x+1}{e^x-1} - 4\right).
\end{equation}
The constants $m_e$, $\sigma_T$, and $c$ are the electron mass, the Thomson scattering cross section, and the speed of light. The angular diameter distance at the source redshift is denoted $D_A(z)$. 

Following the work in \citet{cric16} (hereafter C16), we model the integrated thermal pressure in terms of the energy output from the quasar by writing it in terms of the bolometric luminosity of the quasar such that,
\begin{equation}
    E_{\rm th} = fL_{\rm bol}\tau = \frac{3}{2}\left(\frac{\mu_e}{\mu}\right)\int p_e dV.
    \label{eq:Etherm}
\end{equation}
In the first equality, the total thermal energy scales as the bolometric luminosity over the quasar lifetime $\tau$ times the efficiency with which the radiative energy couples to the surrounding gas, $f$. 
The second equality relates the volume-integrated pressure of the electron gas to the total thermal energy of the gas, where assuming solar abundances and a fully ionized gas the total mean weight per particle is $\mu=0.613$ and the mean weight per free election $\mu_e = 1.17$.
We parameterize the spectral contribution from the SZ effect in our model SEDs in terms of the bolometric luminosity of the quasars in each redshift bin and fit for the coupling factor $f$. 
Because the efficiency is degenerate with $\tau$, we follow the methods of C16 and use a fiducial active quasar timescale. Quasar lifetimes are poorly known, but we use $\tau = \tau_8 \times 10^8$ years, as is commonly suggested by galaxy formation models (\citealt{hopk05b, hopk05c}).
We can ignore any dependence of the cooling time as it is on the order of a Hubble time for the post-shocked gas.
We fit for the coupling efficiency and report it in units of $\tau_8^{-1}$ per cent. 
This parameterization attributes the thermal energy budget of the SZ effect to the feedback energies of the quasars. 

\begin{figure*}
\includegraphics[width=7in]{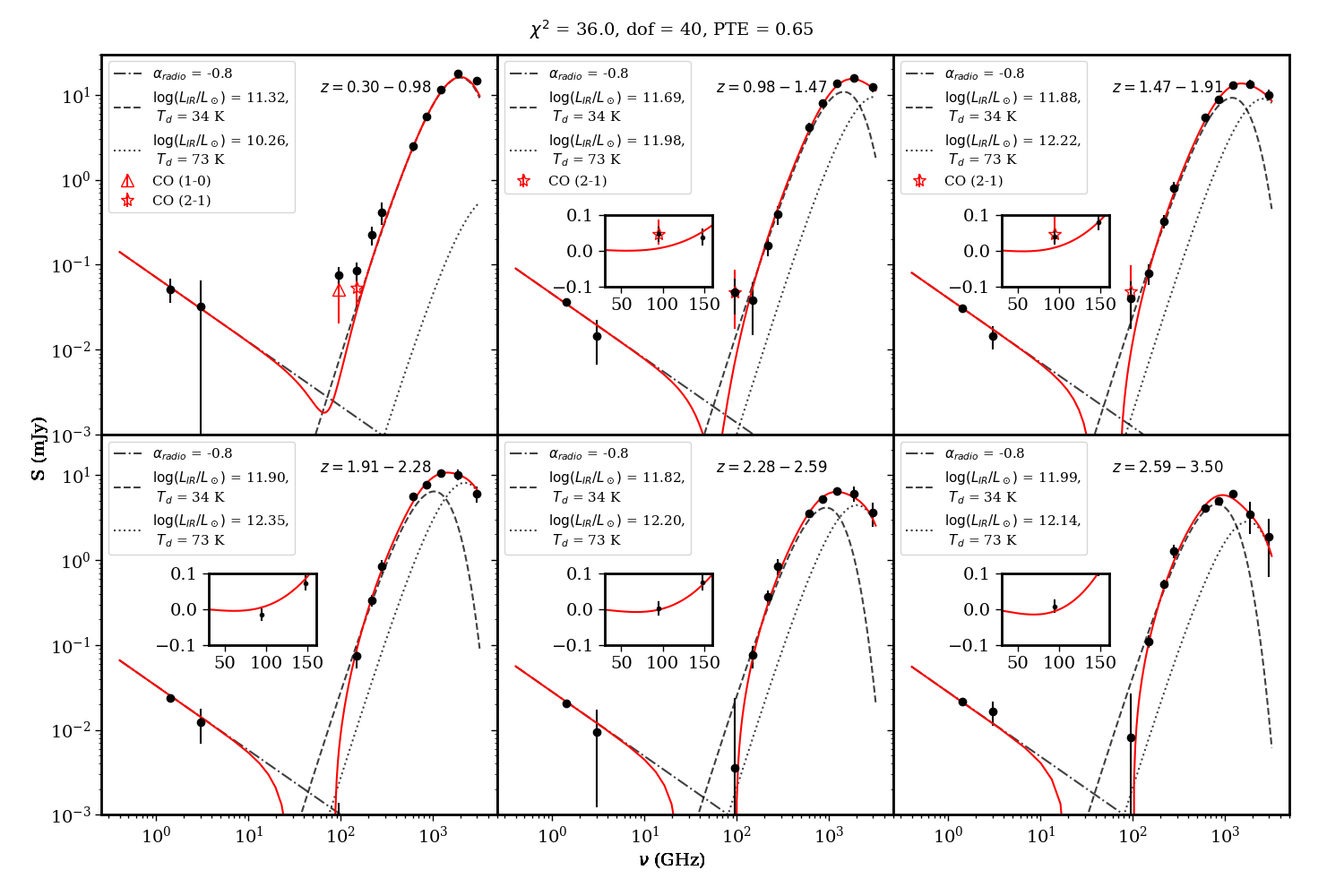}
\caption{Baseline model for the radio through IR SEDs of optically selected quasars. The model is composed of a synchrotron spectrum (dot-dashed line), a two temperature modified blackbody dust spectrum with a cold component (dashed line) and a warm component (dotted line), and a spectral distortion caused by the SZ effect, which results in the decrement in the model at frequencies below 220~GHz. This decrement causes the complete model (red line) to be negative in this frequency regime. The insets display linear plots of the complete model and data in the frequency range from 30-160 GHz for which the complete model becomes negative due to the tSZ spectral distortion. The marginalized amplitudes for the contributions of CO emission lines to the flux density are plotted as red open symbols. The contribution from the CO(1-0) line is plotted as an open triangle at 95~GHz in the first redshift bin. The contributions from the CO(2-1) line are plotted as open stars at 148~GHz in the first redshift bin and 95~GHz in the second and third redshift bins. The complete model and corresponding statistics are the result of fitting all 66 data points simultaneously with the 26 parameter model.  \label{SEDs}}
\end{figure*}

\begin{figure*}
\includegraphics[width=7in]{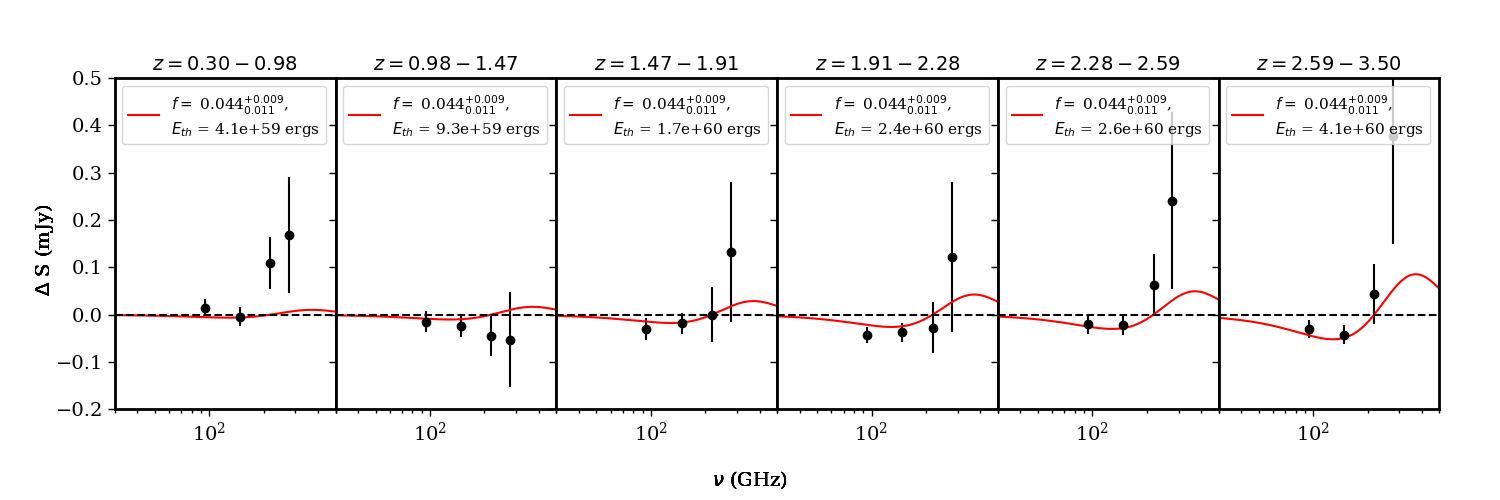}
\caption{Residuals are a result of subtracting synchrotron + dust emission model from the 95~GHz, 148~GHz, 218~GHz, and 277~GHz data points in Figure~\ref{SEDs}. We also subtract the four individual CO emission line contributions to the flux densities at 95~GHz in the first three redshift bins and from at 148~GHz in the first redshift bin. The red curve depicts the SZ spectral distortion using our marginalized value for the coupling fraction when fitting all six redshift bins together $f = 4.4^{+0.9}_{-1.1} \tau_8^{-1}$ per cent. The final constraints on the SZ amplitude are taken from fits to only the three highest redshift bins as discussed in Section~\ref{discussfb}. \label{SZ}}
\end{figure*}

Figure 5 shows that we see excess emission in the 95~GHz and 148~GHz bands in the low redshift bins. 
We test whether the excess flux is due to CO line emission that enters the ACT frequency bands.
The contribution of CO line emission to each band is dependent on the exact shape of each respective band's transmission.
We approximate the bands as top-hat functions and determine the redshift ranges over which we expect the CO emission to enter the bands.
The 95~GHz ACTPol band has a bandwidth of $\sim$35~GHz centered on 94.6~GHz. 
The CO \textit{J}(1-0) line emitting at rest frequency 115~GHz enters this band approximately between $z=0.03$ and $z=0.5$. 
The CO \textit{J}(2-1) line emits at rest frequency 230~GHz. 
This line enters the 95~GHz band between approximately $z=1$ and $z=2$.
The 148~GHz ACTPol detectors have a bandwidths that are 51~GHz (first two detectors) and 41~GHz (third detector), while the ACT equatorial 148~GHz MBAC has a bandwidth of $\sim$18~GHz (\citealt{swet11, thor16}). 
In the lowest redshift bin, this makes the ACTPol 148~GHz data sensitive to the CO \textit{J}(2-1) line emission between $z=0.35-0.87$, and the ACT equatorial 148~GHz data are sensitive to it between $z=0.45-0.65$.
The excess emission is detectable above what can be accounted for by continuum dust emission alone.
We therefore add an additional fitting parameter in the first three redshift bins to fit the amplitude of the CO contribution to the 95~GHz flux density, and in the first redshift bin we add an additional amplitude to contribute to the 148~GHz flux density.
We model and discuss other possible causes for the observed excess in Section~\ref{excess}.
Furthermore, we recognize that higher energy CO emission lines will enter the other two ACT equatorial frequency bands (218 and 277~GHz), as well as continue to shift into all four bands at higher redshifts.
We explore a model for fitting the contribution of the CO lines in all of the ACT stacked data points using a CO spectral line energy distribution (SLED) in Section~\ref{COSLED}. 

We compute the complete model SED for each quasar in the sample, then fit the stacked flux densities with an inverse variance-weighted stacked model. 
We use the same inverse variance weights of the data to construct the model stacked SED,
\begin{equation}
    \bar{S_{\nu}^{\delta}} = \frac{\sum\limits_{i}^{N_{\delta}}w_{i,\nu}S_{tot,i}(\nu,z_i,\alpha,A_{s}^\delta, A_{CO}^\delta, f,\beta, L_{\rm IR,c}^\delta, T_{d,c}, L_{\rm IR,w}^\delta, T_{d,w})}{\sum\limits_{i}^{N_{\delta}}w_{i,\nu}}.
    \label{eq:stackedSED}
\end{equation}
Here, $S_{tot,i}$ is the sum of the synchrotron, CO line emission contribution to the flux, SZ, and dust spectral components for each quasar at redshift $z_i$.
In the base model, there are four total $A_{CO}$ parameters: one each entering the 95~GHz and 148~GHz in the lowest redshift bin, and one entering the 95~GHz band in redshift bins two and three.
The coupling fraction $f$ is a measured quantity and is used to compute $\int p_edV_i$ as a function of $L_{\rm bol,i}$ using Equation~\ref{eq:Etherm}.
The bolometric luminosity is optically derived using the SDSS $i$-band absolute magnitudes, M$_i$($z~=~2$), which are converted to rest frame luminosities $L_{2500}$ at 2500 $\si{\angstrom}$ and then to $L_{\rm bol}$ by multiplying by a bolometric correction of 5 \citep{rich06}. The bolometric luminosity correction of $L_{2500}$ is uncertain by up to 40 per cent, which we return to later in Section ~\ref{discussfb} when discussing the interpretation of our results. 

\subsection{Our Baseline Model Fit Results}

The baseline model is made up of a synchrotron spectrum, a spectral distortion from the SZ effect, four CO-line flux density amplitudes, and two modified blackbodies to describe the dust spectrum. 
We fit all six redshift bins simultaneously.
The radio emission is modelled as an optically thin synchrotron spectrum for which we fit a single spectral index $\alpha$ across all redshift bins and six individual normalization factors, one for each redshift bin. 
The spectral distortion due to the SZ effect is given by Equations~\ref{eq:Ssz} and \ref{eq:Etherm}.
We model the spectral distortion with one value of the coupling fraction at all redshifts in units of $\tau_8^{-1} \mathrm{per cent}$.
In Sections \ref{discussfb} and \ref{comparefb}, we discuss the interpretation of the total measured thermal energy from the SZ signal and the implications for quasar feedback. 
Until then, our discussion of the measured coupling fraction $f$ in each model iteration assumes the total signal can be attributed to quasar feedback energy.
The CO-line flux density model includes three amplitudes to fit the 95~GHz excess in redshift bins 1--3, and one amplitude to fit the 148~GHz excess in redshift bin 1. 
For the two modified blackbody dust spectra, we fix the emissivity to $\beta = 1.5$ and fit for two $L_{\rm IR}$ parameters for each redshift bin and two dust temperatures fixed across redshift bins.
We explore the dependencies of our results on each component of the baseline model in Section~\ref{dependencies}. 

We use a Markov Chain Monte Carlo (MCMC) analysis that steps through the parameter space of the model defined by Equation \ref{eq:stackedSED} plus the four CO amplitudes and computes a Gaussian likelihood function at each step in the chain. 
We use the MCMC ensemble sampler algorithm "emcee" \citep{fore13}.
All of the parameters for which we are fitting are defined with uniform priors. 
The synchrotron amplitudes, CO flux contribution amplitudes, and IR luminosities are constrained to be greater than zero. 
The radio spectral index is set to be between $-1.5\leq \alpha \leq -0.3$. 
The cold and warm dust temperatures are constrained to be less than and greater than 45~K, respectively, and the coupling fraction can vary between $0\leq f \leq 1$. 
We marginalize over the 26 parameters in our baseline model and report the 50th percentiles of the posterior distributions for a subset of these parameters in Table \ref{results}. 
The reported uncertainties are statistical uncertainties given by the 16th and 84th percentiles in the distributions.
Not included in this table are the amplitudes of the synchrotron emission in each redshift bin nor the CO amplitudes. 
To confirm that the chains have sufficiently converged, we inspect each of the parameter trace plots for all walkers in the chain.
These physical quantities are further discussed in Section~\ref{discussLIRLCO}. 
In addition to the marginalized parameters, we report the total thermal energy in each redshift bin calculated from the coupling fraction $f$ in column 4.
We also compute the total far infrared luminosity by summing the two modified blackbodies and integrating those sums from rest frame 8-1000~$\mu$m.
Following \citet{kirk12}, we also compute an effective temperature $T_{d,eff}$ by weighting the dust temperature of the cold and warm components by the fraction of  $L_{\rm IR,tot}$ that is from $L_{\rm IR,c}$ and $L_{\rm IR,w}$, respectively. 
Our measurements of $L_{\rm IR}$ and $T_{d}$ agree well with the \citet{kirk12} measurements, which shows that the quasar dust temperatures are on average $\sim$20 K greater than those for star forming galaxies. 
Our effective temperatures with their statistical uncertainties lie within the range of 350~$\mu$m-derived dust temperatures for high redshift radio quiet quasars from \citet{beel06}.

We find that an SZ spectral distortion in the model SEDs is preferred at the level of 4$\sigma$. 
The best-fitting values at the location of the maximum likelihood in the Markov chains produce a $\chi^2 = 36.0$ for 40 degrees of freedom.
All reported $\chi^2$ values are calculated as $-2 \mathrm{ln} \mathscr{L} _\mathrm{max}$, where ln$\mathscr{L} _\mathrm{max}$ is the value of the log likelihood at the location of the best-fitting parameters (which are, in general, not the same as the reported medians of marginalized parameter distributions).
The amplitude of the SZ effect implies that the radiative energy of the quasar winds couples to the surrounding gas with an efficiency of $f = (4.4^{+0.9}_{-1.1}) \tau_8^{-1}$ per cent. 
Using the bolometric luminosities in each redshift bin and Equation~\ref{eq:Etherm}, we can translate the efficiency into  total thermal energy of the quasar wind that produces the SZ effect.
This quantity is also given in Table~\ref{results} for each redshift bin. 

Figure \ref{SEDs} shows the full radio through far-infrared SEDs of radio quiet quasars in six redshift bins between $0.3 < z < 3.5$. 
The red line shows the complete model given by our marginalized parameters to describe the synchrotron and dust emission plus the spectral distortion from the SZ effect. The SZ effect decrement causes the complete model to become negative between $\sim$10-95~GHz, depending on the amplitude of the SZ signal in a given redshift bin. 
In Figure \ref{SZ}, we plot the residuals of the four ACT data points minus the synchrotron emission, the CO emission flux contribution amplitudes, and the dust spectra for all six redshift bins. 
Plotted in red is the SZ effect spectral distortion as defined by our marginalized value for the coupling fraction when fitting all six redshift bins simultaneously. 
The coupling fraction is converted to the SZ amplitude using Equations~\ref{eq:Ssz} and ~\ref{eq:Etherm}.

\section{Modelling Dependencies}
\label{dependencies}

\subsection{Modified blackbody parameters}
\label{dust}

Our base model uses two modified blackbodies to describe the thermal dust emission with one colder component and one warmer component. 
A two temperature modified blackbody dust emission model is supported by SED fitting of quasars (\citealt{bian19, kirk12, kirk15}) and $\textit{Herschel}$ starburst galaxies \citep{pear13}. 
To explore variations on the base model, firstly we implement a version of the model in which $\beta$ is a free parameter. 
The resulting marginalized $\beta$ is $1.4^{+0.1}_{-0.1}$, with the maximum likelihood corresponding to a $\chi^2$ of 38.5 for 39 degrees of freedom. The resulting coupling fraction from the SZ amplitude is $f = (5.7^{+1.2}_{-1.3})~\tau_8^{-1}$ per cent. 
This is 1.4$\sigma$ greater than our baseline model $f$ in which $\beta$ is fixed to 1.5. 
Because the values of $\beta$ and $f$ are correlated and we wish to compare the baseline $f$ to different model variations, we keep $\beta$ fixed to 1.5 in all model variations unless otherwise specified. 
To further investigate the dependence of $f$ on $\beta$ in each stand alone model variation for which $\beta$ is fixed, we run a version in which we allow $\beta$ as a free parameter and find it ranges between 1.4 and 1.6, so 1.5 is a reasonable fixed value. 
The other dust models that we test include a single temperature, optically thin dust spectrum, and a single temperature, optically thick dust spectrum. 
In each of these tests, the other model components remain the same unless otherwise specified. The results of all of these tests are summarized in Table~\ref{compare}.

In the case of a single temperature, modified blackbody dust spectrum, for which the model allows a greater number of degrees of freedom, we allow $\beta$ to be a free parameter. 
We find that a single temperature, optically thin dust model is not sufficient to describe the data. 
Using this dust spectrum, the best-fitting model produces a $\chi^2$ is 115.7 for 46 degrees of freedom. 
The large $\chi^2$ is coming primarily from the PACS 1875 and 3000~GHz (160 and 100~$\micron$) data points. 
If we exclude these data points in each redshift bin and re-run the fit to the data with the optically thin dust model, the $\chi^2$ statistic becomes 40.3 for 34 degrees of freedom.
This corresponds to a PTE of 0.21, which indicates that this model would be sufficient to describe the lower frequency portion of the dust spectrum.
The coupling fraction measured from this alternative model excluding the PACS data points is $f = $(4.7 $\pm$ 1.2) $\tau_8^{-1} \mathrm{per cent}$, consistent in both value and significance to our base model. 
However, there is no reason to exclude the PACS data in our dust spectrum as they provide valuable information about the shape of the dust emission.
In particular, the PACS data best constrain the warmer component of the two modified blackbody spectra.
These data clearly indicate that an optically thin, single temperature, modified blackbody does not fit our broad wavelength coverage of the far-infrared emission of average quasar SEDs. 

An alternative to a modified blackbody with optically thin dust is a single temperature, modified blackbody spectrum with optically thick dust. \citet{su17}, for example, found this alternative to be a good fit to the SEDs of high redshift dusty star forming galaxies (though this study did not include the higher frequency PACS data). 
The optically thick dust model fit to our SEDs produces a spectrum with an optical depth of $\tau_d=4.3^{+0.5}_{-0.4}$, a dust temperature of ${(74 \pm 3)}$~K, and a feedback coupling fraction of $f= (5.0^{+1.1}_{-1.2})\tau^{-1}_8$ per cent; this model yields a $\chi^2$ of 63.9 for 46 degrees of freedom. 
The PTE for this model is 0.04 and therefore is considered less adequate to fit the dataset as compared to the two modified blackbody dust model. 

\subsection{Radio emission spectrum}
\label{synch}

The origin of radio emission from the hosts of radio quiet quasars remains unknown (\citealt{laor08,zaka16, hwan18, laor19,pane19}). 
Emission in this regime may be due to optically thin synchrotron and free-free emission from star formation in the host galaxy (e.g., \citealt{cond13}), synchrotron and free-free emission from quasar-driven outflows (e.g., \citealt{fauc12b, zubo12, zaka14, nims15}), or synchrotron from a low power jet (\citealt{falc04,leip06}). 
There is also the possibility that there is optically thick synchrotron emission that might be explained, for example, by coronal emission from an accretion disk \citep{laor19}.
In order to better understand the radio emission from radio quiet quasars, we need a large, statistical sample of quasars observed in many frequency bands across the radio regime, which is difficult because the radio emission is weak. 

Stacking analysis can be a powerful tool in addressing this long-standing problem. In our base model, we use an optically thin synchrotron spectrum and keep the spectral index as a fitting parameter. We find a marginalized value of $\alpha = -0.8 \pm 0.2$. 
Our average flux densities at 1.4~GHz and 3~GHz are well fit by this model, and the value of $\alpha$ is indicative of optically thin synchrotron emission, consistent with the well-characterized radio emission from normal star-forming galaxies \citep{cond92}.
We find that there is not a significant contribution to the higher frequency data due to the synchrotron emission, and that a flatter synchrotron slope does not account for the excess millimeter-wavelength emission at low redshift.
We further explore the possible radio emission mechanisms in relation to our observed low redshift excess emission at 95 and 148~GHz in the following section.

Synchrotron emission can also exhibit an exponential cut-off at high frequencies. 
Curved spectral shapes that decline toward increasing radio frequencies have been observed in the spectra of many star forming galaxies (e.g., \citealt{will10, marv15, klei18}). 
One explanation of the decline is that there is an exponential cutoff in the energy spectrum of relativistic electrons at a specific Lorentz factor (\citealt{kard62,jaff73}).
In this scenario, the highest-energy electrons lose their energy much more quickly than the lower-energy electrons producing an abrupt cutoff of the synchrotron spectrum that can be described as $S\propto \nu^\alpha \times e^{-\nu/\nu_b}$. 
The frequency $\nu_b$ at which the spectrum declines  depends on the conditions of the magnetic field that determined the acceleration of the particles, their energy loss, and their escape rates \citep{schl84}. 
\citet{klei18} found cutoff frequencies ranging from 4.7-12.4~GHz.  
Based on these results, we test a synchrotron model with an exponential cutoff frequency of 10~GHz. 
In this model, we fix the spectral slope to the marginalized value of our base model, $\alpha=-0.8$.
This model prefers an SZ spectral distortion at the 4$\sigma$ level with a $\chi^2$ equal to 36.8 for 41 degrees of freedom and a PTE of 0.66.
The marginalized coupling fraction from the SZ amplitude is $(4.1 \pm 0.9)\tau_8^{-1}$ per cent. 
Only half of star-forming galaxies show the exponential cutoff in their synchrotron spectra \citep{klei18}, and even less is known about the synchrotron cutoff of radio-quiet quasars. 
If the synchrotron exponential cutoff frequency is as high as 40~GHz, the SZ amplitude is still measured at the 3.6$\sigma$ level.
Therefore, we do not use the model with a synchrotron cutoff as our baseline model. Rather, we explore this possibility and find that the derived SZ coupling efficiency in the presence of the cutoff at 10~GHz is 1$\sigma$ below that from our baseline model.

\begin{table}
\centering
\caption{Comparison of modifications to our baseline model to describe the SEDs of radio quiet quasars. All variations have $\beta_\mathrm{fixed} = 1.5$ except for the optically thin dust model. The 1$\sigma$ uncertainties are statistical and determined from the 16th and 84th percentiles of the posterior distribution. \label{compare}}
\begin{tabular}{lcccc}
\hline
 Model                        & $f (\tau_8^{-1} \%)$  & $\chi^{2}$ & d.o.f. & PTE    \\
 \hline
                              &                       &            &        &        \\
 Baseline: Two modified BB    & 4.4$^{+0.9}_{-1.1}$   & 36.0       & 40     & 0.65   \\
 dust + indiv. CO Amps        &                       &            &        &        \\
                              &                       &            &        &        \\
 Optically thin dust          & 4.8$^{+1.9}_{-2.1}$   & 115.7      & 46     & 6e-08  \\
 (No PACS)                    & (4.7$^{+1.2}_{-1.2}$) & (40.3)     & (34)   & (0.21) \\
                              &                       &            &        &        \\
 Optically thick dust         & 5.0$^{+1.1}_{-1.2}$   & 63.9       & 46     & 0.04   \\
                              &                       &            &        &        \\
 Baseline + free-free         & 5.0$^{+0.6}_{-0.6}$   & 40.3       & 35     & 0.25   \\
                              &                       &            &        &        \\
 Baseline + opt. thick        & 9.9$^{+2.5}_{-2.2}$   & 29.9       & 35     & 0.71   \\
 synchrotron                  &                       &            &        &        \\
                              &                       &            &        &        \\
 Two modified BB dust         & 4.2$^{+1.5}_{-1.4}$   & 49.3       & 38     & 0.10   \\
 + CO SLED                    &                       &            &        &        \\
                              &                       &            &        &        \\
 Baseline with no CO: $z \geq 1.91$ & 5.0$^{+1.2}_{-1.3}$   & 15.6       & 20     & 0.74   \\
 (Primary Result)             &                       &            &        &        \\
                              &                       &            &        &        \\
\hline
\end{tabular}

\end{table}

\subsection{The 95~GHz/148~GHz excess at $z\leq1.91$}
\label{excess}

Our baseline model accounts for the 95/148~GHz excess emission  with contributions from CO lines entering the ACT and ACTPol passbands (see Figure~\ref{SEDs}), but we test several alternative explanations in this section.
The rest frequency of such emission at these redshifts should lie on the low-frequency tail of the dust spectrum, and the presence of any excess here is difficult to explain. 

We tested a model with the addition of thermal free-free emission with $S_{\nu}\propto \nu^{-0.1}$. 
In this model, we fix the synchrotron spectral index to the marginalized baseline value $\alpha=-0.8$. We allow the amplitude of the free-free emission to vary between redshift bins, which produces marginalized amplitudes that are not consistent between bins. 
We first fit the model in which we replace the CO emission amplitudes with the free-free emission as an alternative explanation.
The best-fitting parameters of this model produce ${\chi^2 = 50.8}$ for 39 degrees of freedom, yielding a PTE of 0.09. 
This model provides a marginally good fit to the data, but in particular, it fails to explain the amplitude of the 95~GHz/148~GHz flux densities at $z\leq1.91$ and has the wrong spectral shape to do so. 
We then consider a model in which there is a free-free emission contribution in addition to CO line amplitudes. This model produces a ${\chi^2 = 40.3}$ for 35 degrees of freedom, yielding a PTE of 0.25. 
This is comparable to our baseline model, so does not justify the addition of the free-free component. It also fails to adequately describe the low-frequency radio slope. 

It is possible that this emission may be the result of a self-absorbed synchrotron emission component (in addition to the optically thin synchrotron emission fitted at lower frequencies), which generates a flat or inverted radio spectrum. 
Excess emission at rest frame $\sim$95~GHz has been observed in a sample of radio quiet quasars (\citealt{beha15, beha18}).
An example of what might cause such a spectrum in radio quiet quasars is a putative coronal emission originating within 1~pc around the quasar \citep{inou14, laor19}.
From the 1.4 GHz and 3 GHz fluxes, it is clear that the radio spectra of our quasar SEDs are not flat at these frequencies, but at $z\leq1.91$, our spectra between 3 GHz and 95 GHz indicate a nearly flat, or even rising, spectral slope.
A few spectra with these shapes have been observed in the radio/millimeter regime of the SEDs of nearby radio quiet quasars and Type 1 Seyfert galaxies (e.g. \citealt{barv96, doi16, beha15}), and explaining this has been challenging. 

We test the possibility that self-absorbed synchrotron emission is contributing to the SED in addition to an optically thin synchrotron component by adding a component with a rising spectrum, $S_{\nu} \propto \nu^{0.5}$. 
This curve is an approximation as the exact form depends on the fraction of non-thermal electrons, the spectral index of the power law distribution of non-thermal electrons, the black hole mass, and mass accretion rate (\citealt{ozel00, inou14}). 
There is also a cut off frequency at which the spectrum falls off, which is around rest frame frequency $\nu=1$~THz. We cut off our self-absorbed synchrotron at this frequency since the relative contribution of this compared to the dust emission is negligible.
We also fix the optically thin synchrotron spectral index to $\alpha=-0.8$.
As with the free-free emission, we first consider optically thick synchrotron as an alternative to CO emission in order to fit the excess. 
This model yields a $\chi^2$ of 36.4 for 39 degrees of freedom and a PTE of 0.59. This model does a better job than free-free in terms of accounting for the millimeter excess and thus represents a viable candidate.

We then include optically thick synchrotron in addition to CO emission in order to explain the excess.
This model yields a $\chi^2$ of 29.9 for 35 degrees of freedom and a PTE of 0.71, a modest improvement over CO alone. 
This model requires a larger SZ amplitude than the base model at the level of 4$\sigma$, with $f=9.9^{+2.5}_{-2.2} \tau_8^{-1}$ per cent, but it is difficult to constrain the amplitude of the optically thick spectra, particularly in the higher redshift bins where the mm-wavelength excess is not observed. 
\citet{beha15}, and subsequent publications from their group, found a relationship between the mm-wavelength and X-ray luminosities of a sample of quasars that is consistent with the coronal emission models: $L ~(95 ~\mathrm{GHz})\sim10^{-4}L_X~(2-10~\mathrm{keV})$ . 
To test the validity of our results, we compare the 95~GHz luminosity from the optically thick spectra to $L_X~(2-10~\mathrm{keV})$ estimated from the bolometric luminosity using a correction factor $k_{bol}=12$ \citep{luss10}. 
We find $L ~(95 ~GHz)/L_X~(2-10~\mathrm{keV}) = (5.1, 1.6, 3.7, 5.6, 17, 28)\times10^{-4}$ as a function of increasing redshift bin. 
At low redshift, the ratios are similar to that found by \citet{beha15, beha18}. 
At higher redshifts, our data do not constrain the optically thick component, making the fit values (and associated SZ values) suspect.

We explore the possibility that this excess emission is a result of anomalous microwave emission (AME). AME is best understood as spinning dust emission that contributes to radio emission between rest frame $\sim10-90$~GHz with a peak at 30~GHz \citep{dick18}. First discovered as a part of Galactic foregrounds in cosmic microwave background surveys, it has recently been detected in two extragalactic sources (\citealt{hens15, murp18}). AME as the origin of the observed excess in our dataset is likely ruled out because the excess we observe is at rest frame frequencies $\gtrsim$123~GHz, and while a slight excess is expected in the high frequency tail of the anomalous emission spectrum, the observed excess is difficult to reconcile with standard AME models. To estimate an expected AME flux density at 30~GHz, we use the relationship between dust radiance, or $F_{\rm IR}$, and the 30~GHz AME flux density amplitude within the Galaxy from \citet{hens16}. We arrive at a value of expected $S_{30~GHz}^{AME}\sim10$~$\mu$Jy in our lowest redshift bin using the measured total infrared flux from our model. \citet{murp18} find their measured AME at 30~GHz to be a factor of two greater than that expected from the Galactic relation in NGC 4725. This puts the maximum expected peak AME flux at $\sim20$~$\mu$Jy, and the spinning dust models exhibit a steep drop at frequencies higher than the peak. Our measured 95~GHz excess in the lowest redshift bin is $\sim70$~$\mu$Jy after subtracting the baseline dust model, which is already a factor of 3 greater than generous estimate of the expected peak flux from AME. We therefore conclude that AME is unlikely to produce such a strong excess at our observed frequencies.  


\subsection{CO Spectral Line Energy Distribution}
\label{COSLED}

In our base model, we include additional amplitudes to account for the CO(1-0) and CO(2-1) emission lines entering the ACT(Pol) 95~GHz and 148~GHz bands in the first redshift bin, and to account for the CO(2-1) emission line entering the ACTPol 95~GHz band in the second and third redshift bins. 
As the spectrum rises toward the thermal peak, higher $J$ transitions of CO make progressively smaller fractional contributions to the observed SED. 
Although we do not detect the higher $J$ transitions above the dust spectrum, we implement an alternative model to test the effect of any flux from the higher $J$ transitions entering the ACT bands.

We thus focus on the CO emission line contributions to the four ACT bands, and only include contributions from energy lines up to CO(5-4). 
This choice is well motivated because CO lines in this energy regime have been detected in quasars out to $z\sim6$. 
We use the CO line luminosity ratios given in of Table 2 of \citet{cari13} to define the spectral line energy distribution (SLED) expected for quasars up to the CO(5-4) transition. 
For each quasar at redshift $z$, we calculate which CO lines enter into which frequency bands and allow a contribution to the flux density from CO emission lines in those bands.
We then fit for the amplitude of the CO(1-0) line luminosity in each redshift bin, using the luminosity ratios as dictated by the SLED to account for the relative amplitude of the higher energy line transitions.
For redshift bins 1 and 2, this includes a CO emission component in all four ACT bands, and for redshift bins 3 and above this includes CO emission line components in only the 95 and 148~GHz bands. 
We refer to this as the CO SLED model.
This model produces a $\chi^2$ of 49.3 for 38 degrees of freedom and a PTE of 0.10. 
The increase in $\chi^2$ above that of the baseline model is driven by the 95/148~GHz excess. We further discuss the implications for the CO emission in the next section.
The resulting coupling fraction of $f=(4.2^{+1.5}_{-1.4})\tau_8^{-1}$ per cent is compared with the other model iterations in Table~\ref{compare}.



\section{Discussion}

Here, we discuss the physical validity and implications of our SED fitting. We describe the resulting dust spectra and CO fluxes in \ref{discussLIRLCO}, and in \ref{discussfb} we return to our task of estimating the contribution of quasar feedback to the measured total thermal energies. We further compare to other measurements in the literature of the thermal SZ effect in quasars host halos in \ref{comparefb}. 

\subsection{Far-infrared and CO luminosities}
\label{discussLIRLCO}

 \begin{figure}
\includegraphics[width=3.4in]{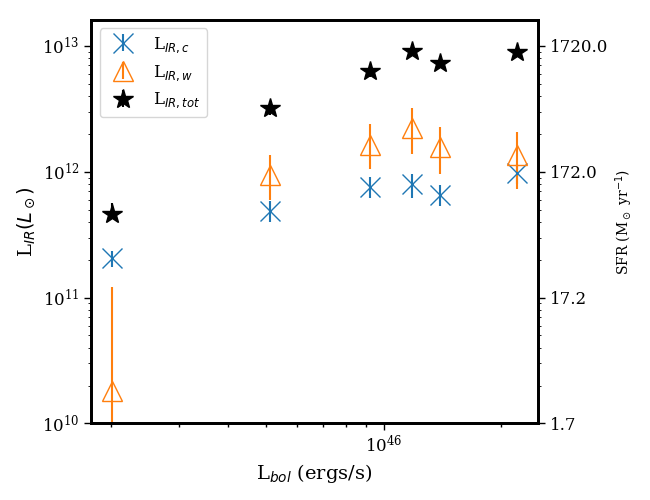}
\caption{Infrared luminosities of the warm (orange triangles) and cold (blue X's) components of our baseline model SEDs, and the total infrared luminosities (black stars) computed from the sum of the two modified blackbody components as a function of bolometric luminosity determined from $L_{2500}$. The second y-axis gives the upper limit on star formation rate calculated from the IR luminosities. \label{LIR_Lbol}}
\end{figure}

\begin{figure}
\includegraphics[width=3.25in]{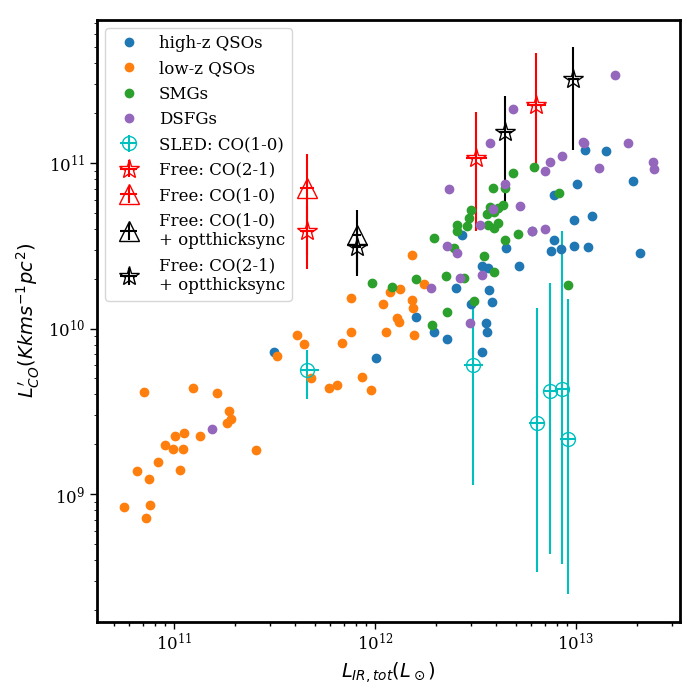}
\caption{$L'_{CO}$ as a function of total infrared luminosity in comparison with low redshift ($z\lesssim$ 0.4) quasars (orange, \citealt{evan06, xia12}), high redshift ($z\gtrsim$ 1) quasars (blue, \citealt{walt03, solo05, riec06, cari07, maio07, copp08, wang10, simp12}), sub-millimeter galaxies (green, \citealt{grev05, fray08, dadd09, ivis11, both13}), and dusty star-forming galaxies (purple, \citealt{grev14}). The $L'_{CO}$ values that are obtained by directly fitting the CO(1-0) emission contribution to the flux densities in the SEDs are plotted as triangles. The $L'_{CO}$ values that are obtained by fitting the CO(2-1) emission contribution to the flux densities in the SEDs and then converted to the CO(1-0) line luminosity are plotted as stars. Red stars/triangles are from the results of our baseline model (see Figure~\ref{SEDs}), and black points are from the results of adding the optically thick synchrotron component. The cyan open circles are the results from fitting our CO SLED model. This plot highlights the difficulty in explaining the 95/148~GHz excess as our individual CO amplitudes (red and black) produce more CO emission than expected in quasar hosts, while the CO SLED model under produces this emission. \label{LIR_LCO}}
\end{figure}

We compute the total infrared luminosity by integrating the sum of the two modified blackbodies from rest frame 8-1000~$\micron$ in each redshift bin. 
The marginalized warm and cold $L_{\rm IR}$'s from the two temperature modified blackbody dust spectrum are plotted in Figure~\ref{LIR_Lbol} along with the total infrared luminosities as a function of median bolometric luminosity derived from $L_{2500}$. 
A modified blackbody can underestimate the total infrared luminosity of an AGN due to an additional power law component of the AGN continuum spectrum below $\sim$50$\micron$. We use an AGN template from \citet{kirk12} to estimate the additional contribution to the total IR luminosity from this component below 50$\micron$ and find that it could increase the total $L_{\rm IR}$ by up to 15\%. This percent contribution corresponds to 0.6$\sigma$ in the lowest redshift bin and 1.8$\sigma$ in the highest redshift bin.  
 
Disentangling the total infrared emission that is a result of star formation versus that which is coming from the active supermassive black hole accretion is a difficult problem.
\citet{kirk12} perform a detailed SED decomposition of star-forming galaxies, galaxies with active galactic nuclei, and composites.
\citet{kirk12} find  at $z\sim1$ that 21 per cent of the infrared emission can be attributed to star formation, while at $z\sim$2 the contribution increases to 56 per cent.
Furthermore, at the resolution of the ACT and \textit{Herschel} beams, we expect some contribution from clustered dusty star-forming galaxies to the total infrared luminosity.  
The relative contribution of infrared emission from clustered sources will furthermore depend on instrument resolution.
We do not have full decomposition capabilities in this study, but we use the $L_{\rm IR}$-star formation rate relation from \cite{bell03} to compute upper limits on the amount of star formation contributing to our average SEDs.
In Figure~\ref{LIR_Lbol}, we plot a second y-axis that is star formation rate in solar masses per year.
If the entirety of the infrared luminosity is due to star formation in the host galaxies of the average quasar SEDs, then at $z\sim0.7$ the maximum star formation rate is $70~\mathrm{M}_\odot\mathrm{yr}^{-1}$, and at $z\ge2$ it is approximately $790~\mathrm{M}_\odot\mathrm{yr}^{-1}$. 
A more reasonable estimate might be to assume that the cooler component of the two modified blackbody spectrum is due to dust emission heated by star formation.
In this case, the star formation rates range from $36~\mathrm{M}_\odot\mathrm{yr}^{-1}$ to $168~\mathrm{M}_\odot\mathrm{yr}^{-1}$. 
These lower star formation rates agree well with those from \citet{kirk12}. 

From the marginalized CO amplitudes, we compute CO luminosities by first converting the measured flux densities to Watts per square meter using the associated millimeter bandwidth.
We then compute $L'_{CO}$ using the relation:
\begin{equation}
    L^\prime_{\rm CO} = 9.8\times 10^{44} S^\prime_{\rm CO} \frac{ (1+z) D_{\rm A}^2}{\nu^3_{\rm obs}}\,  \rm K\, km\, s^{-1}\, pc^2,
\end{equation}
where $S^\prime_\mathrm{CO}$ is the flux in units of W/m$^2$, $\nu_{obs}$ is the observed frequency in units of Hz of the CO emission line at a given redshift, and $D_A$ is the angular diameter distance in pc \citep{cari13}. 
We compare our derived $L'_{CO}$ values with expectations for quasars at low redshift ($z\lesssim$ 0.4, \citealt{evan06, xia12}) and high redshift ($z\gtrsim$ 1, \citealt{walt03, solo05, riec06, cari07, maio07, copp08, wang10, simp12}), submillimeter galaxies (\citealt{grev05, fray08, dadd09, ivis11, both13}), and dusty star-forming galaxies \citep{grev14}. 
All of these line luminosities have been corrected to represent $L'_{CO}$(1-0) (see \citealt{ryle19} for more detail on the quasar and sub-mm literature values). 
We correct our $L'_{CO}$ values that are derived from higher energy transitions using the \citet{cari13} line ratios. 

We plot $L'_{CO}(1-0)$ vs. $L_{\rm IR,tot}$ in Figure~\ref{LIR_LCO} for our base model where in each redshift bin in the range $z=0.3-1.91$ we fit for CO(1-0) and CO(2-1) amplitudes independently (open red symbols) and for the model in which we add the optically thick synchrotron spectrum to our base model (open black symbols). 
\citet{kirk19} show that the relationship between the CO line luminosity and IR luminosity in active galaxies should be with respect to only the IR luminosity due to star formation. If we were to plot just the $L_{\rm IR}$ due to star formation, all of the quasar points including ours would shift to the left in Figure~\ref{LIR_LCO}. The literature values of infrared luminosity obtained for this comparison are total $L_{\rm IR}$, so we keep the same convention for our derived values.

In Figure~\ref{LIR_LCO}, we also show the results from our CO SLED model in which we fit for one CO(1-0) amplitude in each redshift bin (cyan open circles) with the higher $J$ line amplitudes correspondingly constrained by the SLED. 
The triangle points are derived directly from the marginalized CO(1-0) amplitudes, while the stars are the independently fitted CO(2-1) amplitudes that have been corrected to $L'_{CO}(1-0)$ using the ratio $L'_{CO}(1-0)$/$L'_{CO}(2-1) = 0.99$ from \citet{cari13}.
In our base model with independently fitted CO amplitudes, our marginalized flux densities correspond to $L'_{CO}(1-0)$ line luminosities that are $\sim$1$\sigma$-2$\sigma$ larger than the literature values. 
While the CO amplitudes are in better agreement when adding the optically thick synchrotron spectrum to our base model, they are still biased high by $\sim1\sigma$ in comparison to the literature values.
The CO SLED parameterization yields a marginalized CO(1-0) flux in the lowest redshift bin (cyan circle with the smallest infrared luminosity) that agrees well with the literature values of $L_{\rm IR}$ vs. $L'_{CO}(1-0)$, and all of the higher redshift bins produce $L'_{CO}(1-0)$ line luminosities that are $\lesssim$1$\sigma$ below their expected values given their infrared luminosities.
Furthermore, the CO SLED model does not produce an adequate fit to the 95/148~GHz data points, which is driving the poor $\chi^2$ in this model variation.
The base model with independently fitted CO amplitudes fits the data well, but overestimates the expected CO line luminosities at the derived total infrared luminosities.
This indicates the presence of an additional emission component at observed 95~GHz that is not accounted for in any of our model iterations, such as the optically thick synchrotron addition for which the validity was discussed in Section~\ref{excess}. 

\subsection{Implications for quasar feedback}
\label{discussfb}

\begin{table*}
\centering
\caption{Model results for the three redshift bins at $z\geq$1.91 only, with the marginalized parameters derived from the 50th percentile of the posterior distributions from the MCMC analysis. The fit parameters include the radio spectral index $\alpha$, the coupling fraction of the SZ effect $f (\tau_8^{-1} \mathrm{per cent})$, and the warm and cold components of the two modified blackbody dust spectra parameterized by the infrared luminosities $L_{\rm IR}$ and the dust temperatures $T_d$. The total thermal energy $E_{\rm th}$, total infrared luminosity $\log(L_{\rm IR,tot})$, and effective dust temperature $T_{d,eff}$ columns are calculated from the fit parameters. The 1$\sigma$ statistical uncertainties are given by the 16th and 84th percentiles and propagated accordingly for the calculated values.  \label{results456}}
\begin{tabular}{lccccccccc}
\hline
 $z$ range   & $\alpha$       & $f (\tau_8^{-1} \%)$ & $E_{\rm th}$ (ergs)                     & $\log(L_{\rm IR,c} [L_\odot])$ & $T_{d,c} (K)$  & $\log(L_{\rm IR,w} [L_\odot])$ & $T_{d,w} (K)$  & $\log(L_{\rm IR,tot} [L_\odot])$ & $T_{d,eff}$    \\
 \hline
             &                      &                      &                                     &                            &                &                            &                &                              &                \\
 1.91 - 2.28 & -0.5$^{+0.1}_{-0.2}$ & 5.0$^{+1.2}_{-1.3}$  & (2.7$^{+0.7}_{-0.7})\times10^{+60}$ & 11.7$^{+0.2}_{-0.2}$       & 31$^{+3}_{-3}$ & 12.6$^{+0.1}_{-0.2}$       & 66$^{+7}_{-6}$ & 13.0$^{+0.1}_{-0.1}$         & 62$^{+7}_{-5}$ \\
             &                      &                      &                                     &                            &                &                            &                &                              &                \\
 2.28 - 2.59 &                      &                      & (3.0$^{+0.7}_{-0.8})\times10^{+60}$ & 11.6$^{+0.2}_{-0.2}$       &                & 12.4$^{+0.1}_{-0.2}$       &                & 12.8$^{+0.04}_{-0.1}$         & 61$^{+7}_{-5}$ \\
             &                      &                      &                                     &                            &                &                            &                &                              &                \\
 2.59 - 3.50 &                      &                      & (4.7$^{+1.2}_{-1.2})\times10^{+60}$ & 11.8$^{+0.2}_{-0.2}$       &                & 12.4$^{+0.1}_{-0.2}$       &                & 13.0$^{+0.04}_{-0.1}$         & 60$^{+6}_{-5}$ \\
\hline
\end{tabular}
\end{table*}

\begin{figure*}
\includegraphics[width=7in]{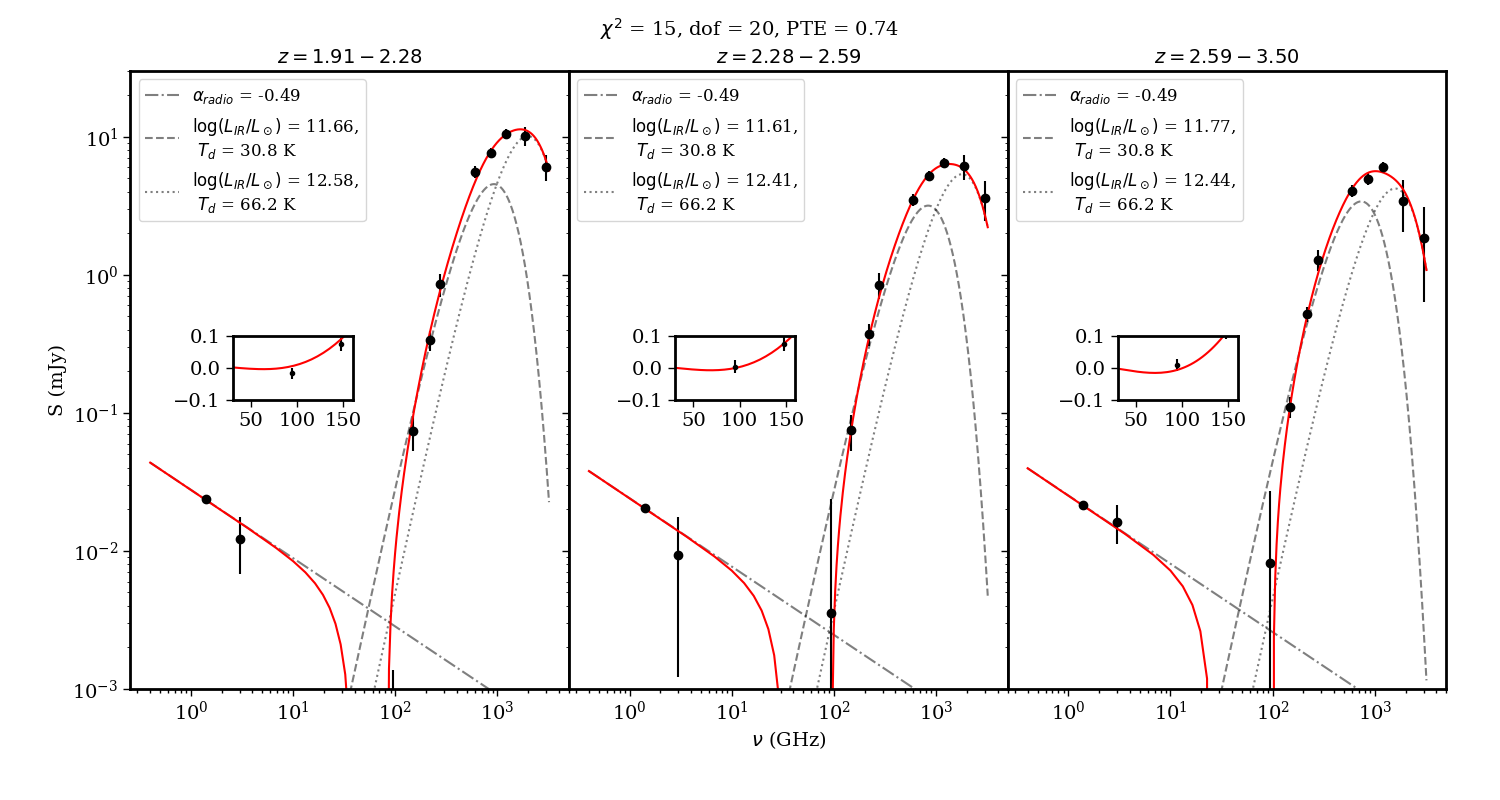}
\caption{Marginalized model for the radio through IR SEDs of optically selected quasars at $z\geq$1.91. The model is composed of a synchrotron spectrum (dot-dashed line), two modified blackbody dust spectra with a cold component (dashed line) and a warm component (dotted line), and a spectral distortion cause by the SZ effect, which results in the decrement in the model at frequencies below 220~GHz. This decrement causes the complete model (red line) to be negative in this frequency regime. The insets display linear plots of the complete model and data in the frequency range from 30-160 GHz where the complete model becomes negative due to the tSZ spectral distortion. The complete model and corresponding statistics are the result of fitting all 33 data points simultaneously with the 13 parameter model. \label{SED456}}
\end{figure*}

\begin{figure*}
\includegraphics[width=7in]{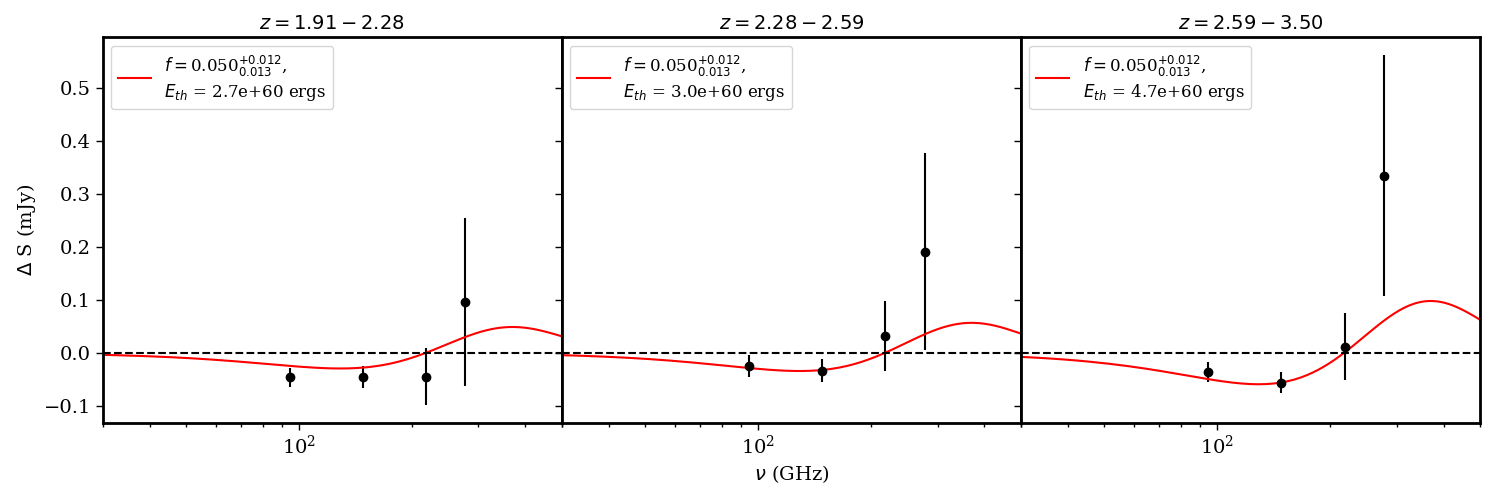}
\caption{Residuals are a result of subtracting the synchrotron + dust emission model from the 95~GHz, 148~GHz, 218~GHz, and 277~GHz data points in the model fit only to the three highest redshift bins, $z\geq$1.91. The red curve depicts the SZ spectral distortion using our marginalized value for the coupling fraction, $f$=(5.0$^{+1.2}_{-1.3}$)$\tau_8^{-1}$ per cent. \label{SZ456}}
\end{figure*}


Due to the uncertainty in determining the emission mechanism causing the excess flux densities at low redshift in the ACT frequency bands, we limit our discussion of the implications for quasar feedback to the three highest redshift bins, $z\geq$1.91. 
These three bins do not exhibit excess flux above any reasonable dust spectrum, and we fit these data with our base model of synchrotron and two modified blackbody dust spectra, plus a spectral distortion from the SZ effect. 
The results of fitting to only the three highest redshift bins are given in Table~\ref{results456}. 
The model fits these data with a $\chi^2$ of 15.6 for 20 degrees of freedom and a PTE of 0.74. Figure~\ref{SED456} shows the marginalized model in red plotted over the data, and Figure~\ref{SZ456} shows the residuals of the data minus the synchrotron and dust spectra with the SZ spectrum shown in red. 
In comparison with the base model that simultaneously fits all six redshift bins, the amplitude of the SZ effect and the coupling fraction exhibit a less than 1$\sigma$ increase, while the parameters to describe the synchrotron and dust spectra change more significantly. 
The spectral index of the synchrotron flattens from $-0.8\pm0.2$ when fitting all six bins to $-0.5^{+0.1}_{-0.2}$. 
Though this is a 1.5$\sigma$ change, there is only a 0.5$\sigma$ increase in the SZ amplitude, indicating that this is not a strong driving factor in measuring the SZ effect. 
The cold components of the dust spectra all exhibit a decrease in amplitude and temperature, and the warm components increase in amplitude and decrease in temperature by $\sim$1$\sigma$.
The key conclusion is that restricting the fitting from 6 redshift bins to 3 redshift bins changes the SZ coupling factor by less than 1$\sigma$.
Furthermore, producing the observed decrement in flux compared to the low frequency tail of the modified blackbody dust spectrum at high redshifts is difficult with anything other than the SZ effect.

For the three highest redshift bins fitted simultaneously, our model of the radio through far-infrared SEDs prefers the addition of the SZ effect component at a significance of 3.8$\sigma$. 
Parameterizing the model such that the amplitude of the SZ effect scales with the bolometric luminosities of the quasars and assuming the entirety of the signal is attributed to the energy output of the quasar, we find that the radiative output of the quasar couples to the surrounding gas with an efficiency of $(5.0^{+1.2}_{-1.3})$  per cent for a fiducial quasar lifetime of 10$^8$ years. 
This value agrees well with those used in simulations of galaxy evolution, which invoke quasar feedback with a coupling efficiency of 5-7 per cent in order to reproduce the luminosity function of the observed universe \citep{hopk06}. 
Our assumption linking the amplitude of the SZ effect to the bolometric luminosities of quasars indicates a lower signal for the thermal SZ effect around lower redshift quasars.
This is contrary to what might be expected when considering the detection of fossil feedback energy due to the long cooling times in quasar dark matter halos \citep{plan13,grec15,spac16,spac17,spac18,tani19,pand19}.
However, if we consider that quasars are a phase in galaxies' evolution \citep{Hopk05a}, then the quasars that we observe at lower redshifts may not have been quasars at high redshift. 
Furthermore, some evidence suggests that low redshift quasars are hosted in lower-mass dark matter haloes (e.g. \citealt{rich12}). 
Therefore, the quasar energy injection (and therefore SZ signal) could plausibly be smaller at low redshift.

Additionally, \citet{cunh13} detail the effect of the CMB blackbody spectrum on the ability to measure a galaxy's intrinsic dust spectrum as a function of redshift and frequency. The CMB spectrum contributes more significantly at higher redshift and lower frequency, and we calculate the fraction of the intrinsic dust emission that we can measure at z=2 and z=3 based on the effective dust temperatures computed from our two-temperature blackbody models. At 148 GHz, we can expect to measure 91.5\% and 86.3\% of the intrinsic dust emission at z=2 and z=3, respectively. This implies that the intrinsic dust spectra could have up a 13.7\% increased flux at this frequency, but these are well within our tSZ uncertainties.  

The coupling fraction measured in this study is 65 per cent lower than that reported in C16, who derive a coupling fraction of $f$= (14.5$\pm$ 3.3)$\tau^{-1}_8 \%$ of the bolometric luminosities of the quasars between $0.5\leq z \leq3.5$; however, as pointed out in C16, these values have a systematic uncertainty that could be as large as 40 per cent according to the scatter in the determination of the bolometric luminosity correction of $L_{2500}$. 
We have made a number of improvements to the sample and the model by comparison to C16.
The quasar sample used in this study is $\sim$5 times larger than used in C16, reducing the statistical uncertainties in the stacked data by up to a factor of 2. 
The addition of the 95~GHz ACTPol data provides an additional direct probe of the SZ effect, and requires the addition of the synchrotron emission spectrum at lower frequencies.
This emission component was left out of the C16 study under the assumption that the contribution to the 148~GHz flux density from the synchrotron emission is negligible. 
This present study is also able to better describe the dust emission spectrum due to the addition of the higher frequency \textit{Herschel} data from the PACS instrument at 1875 and 3000~GHz (160 and 100 $\micron$). 
In this way, we find that the two temperature modified blackbody emission model provides a significantly better fit to these data than any of the other dust models.
In C16, the model with two modified blackbody dust spectra was sufficient to fit the data without any SZ signal, but was not preferred due to the added complexity of the model compared to the number of degrees of freedom.
Thus we suggest that our more refined model, incorporating higher frequency PACS results, provides a more plausible estimate of the SZ component.

For a direct comparison to C16, we test the effects of the additional data points used in this study by excluding the 1.4-95~GHz data points and the PACS data points, and limit the 148~GHz data point to only include quasars that lie within the original ACT equatorial 148~GHz map. We fit these SEDs with an optically thin, single temperature dust spectrum plus the spectral distortion from the SZ effect.
We also use the C16 relation between the total thermal energy and the $\int p_e dV$ of the electron gas, which differs from our Equation~\ref{eq:Etherm} by a factor of 1.03. 
The resulting marginalized coupling fraction corresponding to the thermal SZ amplitude is ${f=(12.3 \pm 3.4) \tau^{-1}_8 \%}$, in agreement within 0.6$\sigma$ of the C16 value. 

There is extensive discussion in C16 of the expected total thermal energy in the dark matter halos hosting quasars due to virialization.
We summarize the main points here, and expand upon that discussion with respect to distinguishing the SZ signal contribution from quasar feedback from that of the virialized halo gas. 
\citet{plan13} find a relation between the integrated thermal SZ signal and halo mass for halo masses down to $\sim3\times10^{13}$~M$_\odot$, and C16 extrapolate it down to lower mass halos that host quasars. 
C16 redefine the relation to be with respect to $M_h = M_{200}$, the mass enclosed within the radius where the density is 200 times the critical density, and this is the same halo mass definition we adopt here.
We report here an overview of expected quasar halo masses, and then use Equation 12 of C16 to compute the total thermal energy expected due to virialization in quasar dark matter halos.

Measurements of quasar host dark matter halo masses range from 1$\times$10$^{12}$~M$_\odot$ to $\gtrsim 5\times10^{12}$~M$_\odot$ (\citealt{rich12,sher12,whit12,shen13,dipo14,dipo15,wang15,Efte19, geac19}). 
The sample of optically selected quasars we use in this study from the SDSS-BOSS DR14 quasar catalog have a median redshift of 1.91. 
\citet{Efte19} present halo mass results from a clustering analysis of a sample of quasars drawn from the same parent quasar catalog with a mean redshift of $z\sim1.5$. 
Using a halo occupation distribution model, they find the  minimum halo mass to host a central quasar to be $2.3\times10^{12}~h^{-1}\mathrm{M}_\odot$.
\citet{geac19} also study the halo masses of a sample of quasars with $\langle z \rangle=1.7$ drawn from the same SDSS DR14 catalog using methods independent of the modelling assumptions inherent in clustering measurements.
They measure the deflection of CMB photons caused by dark matter halos hosting quasars and measure a mean halo mass of ${M_h \sim4\times10^{12}h^{-1}\mathrm{M}_\odot}$. 
At somewhat higher redshifts, \citet{whit12} and \citet{wang15} use the SDSS-III quasars in clustering studies with mean redshifts of $\langle z \rangle=2.4$ and $\langle z \rangle=2.5$, respectively. 
These results both find that quasars reside in dark matter halos of masses  $\sim2\times10^{12}~ \mathrm{M}_\odot$. 
\citet{rich12} implements a halo occupation distribution model on the clustering of $\langle z \rangle=1.4$ quasars, and separately on the $\langle z \rangle=3.2$ quasar sample from \citet{shan12}.
For the low and high redshift samples, they derive distributions of halo masses whose medians are ${4.1\times10^{12}~h^{-1}\mathrm{M}_\odot}$ and greater than ${1\times10^{13}~h^{-1}\mathrm{M}_\odot}$, respectively. 
C16 analytically approximate the full halo mass distributions, and find that the median value of the halo mass depends greatly on the high mass end of the quasar halo mass distribution, which is poorly constrained \citep{rich12}. 

To take a conservative estimate of how much the virial gas in the quasar host halos can contribute to the observed thermal SZ effect, we consider haloes with $M_h = 5\times10^{12}~\mathrm{M}_\odot$, which is among the larger of the values suggested in the literature. 
We compute the total thermal energy expected due to virialization of the dark matter halo with this fixed mass at the average redshift $\langle z \rangle=2.25$. 
Using this mass and redshift, the expected $\int p_e dV$ from virialization is 3.6$\times$10$^{59}$ ergs.
This implies a total thermal energy, using Eq.~\ref{eq:Etherm}, of 1$\times$10$^{60}$ ~ergs.
Taking the inverse variance weighted average of the calculated thermal energies at $z>1.91$ in Table \ref{results456}, the total thermal energy from the SZ effect is ${3.1\times10^{60}}$ ~ergs.
We thus conclude that the measured thermal energy from the SZ signal can be attributed to the hot halo gas in virial equilibrium at the level of $(\sim32\pm8)$ per cent at $z\sim2.25$.
Therefore, between redshifts $z\sim2-3$ we measure a $\gtrsim 60$ per cent excess of thermal energy compared to that expected from virialized gas. 
If instead, typical halo masses hosting quasars are at the low mass end of literature measurements, $M_h = 2\times10^{12}~h^{-1}$M$_\odot$, then $\gtrsim$90 per cent of our measured thermal energy from the SZ effect is in excess of that expected from virialization.

Alternatively, we can turn this calculation around and use the measured thermal energy to put an upper limit on the characteristic halo mass hosting quasars.
Let us assume that the total thermal energy implied by our measurement of the thermal SZ effect may be completely explained by hot halo gas in hydrostatic equilibrium. 
Using our value at $z\sim2$ of ${2.7\times10^{60}}$~ergs means that the effective halo mass of quasars would be ${6.3\times10^{12} ~h^{-1}}$M$_\odot$. 
This is a factor of 2.7 greater than the \citet{Efte19} measurement, and a factor of 1.6 greater than the \citet{geac19} measurement that is independent of clustering at $z\sim1.7$. 

An additional consideration when trying to distinguish the thermal energy from quasar feedback and that from the virialized dark matter halo is the distribution of dark matter halo masses hosting quasars. 
The calculation of thermal energy relies on the halo mass raised to the power of $(5/3)$, and $\langle M \rangle^{(5/3)}$ does not equal $\langle M^{(5/3)} \rangle$. 
Using an approximation of the quasar mass distribution from \citet{rich12}, we find that the characteristic mass computed as $\langle M^{(5/3)} \rangle^{(3/5)}$ is greater than $\langle M \rangle$ by a factor of 1.2. 
Therefore, the mass-inferred SZ signal (or thermal energy) will be different by a factor of 1.35.
This factor depends on the exact shape of the quasar halo mass distribution, and future datasets will better serve to disentangle the contributions to the thermal energy from quasar winds and the virialized halo gas at different mass scales. 

\subsection{Comparisons with Planck measurements of the thermal SZ effect in quasar host halos}
\label{comparefb}

The most recent measurements of the SZ in quasar hosts relevant to our work in C16 and this paper are \cite{verd16} and \cite{soer17}. Both of these use data primarily from the \textit{Planck} satellite (with the addition of Akari data in \cite{soer17} to better constrain the peak of the thermal SED). Since the primary difference between these studies and ours is in the choice of millimeter-wavelength dataset (\textit{Planck} vs. ACT), and since the \cite{soer17} analysis uses SED modelling most transparently related to our work, we restrict our comparisons to \cite{soer17}. Much of our discussion is driven by the difference in the angular resolution: $\sim$5-10$'$ for \textit{Planck} and $\sim$1-2$'$ for ACT and \textit{Herschel}.  

Modelling of the radio-to-far-infrared emission is crucial to detecting and measuring the amplitude of the SZ effect. 
Comparisons between studies must account for differences between the datasets used. 
Along these lines \cite{soer17}  draw attention to the differences between dust parameters derived in C16 ($T_{d}\approx40$~K, $\beta\approx1.1$) and their work ($T_{d}\approx25$~K, $\beta\approx2.3$) to call into question the validity of emission modelling and, by extension, constraints on the SZ signal. 
However, this difference is expected between the \textit{Planck} and ACT/\textit{Herschel} data. 
As pointed out by \cite{soer17}, their dust properties are characteristic of the dusty star forming galaxies clustered around the more massive quasar hosts. 
While the ACT and \textit{Herschel} data of C16 and this study do have contributions to the dust spectrum from clustered sources on scales of the one halo term, these datasets also have a higher contribution from the quasar host itself, where higher temperatures and different dust properties are expected \citep{bian19, hall18, kirk12}.

Similarly, it is important to distinguish the contribution to the measured SZ effect due to the quasar host halos from that due to correlated halos. \citet{soer17} reported the energy from SZ to be $5\times10^{60}$ ergs at 3-4$\sigma$ statistical significance, but also with warnings that systematic errors in model choice made the estimate more uncertain. \cite{hill18} have shown that, at \textit{Planck} resolution, the stacked SZ effect measured for halos in the range ${\mathrm{M}_\odot = (1-5)\times10^{12}}$ (i.e., characteristic of quasar hosts) is dominated by hot gas in large-scale correlated structure, and not by the thermal energy of the quasar host itself. 
This effect is also consistent with simulations by \cite{soer17}. 
In their Figure 12, the thermal energies inferred for quasar host halos are an order of magnitude larger than those expected from virialization. 
The authors point out that the scatter in the simulated thermal energy estimates, which is approximately $5\times10^{60}$~ergs, is primarily due to hot gas in the large scale structure within a \textit{Planck} beam. 
Though not pointed out explicitly in that work, this same structure should bias the stacked simulation high, presumably by the observed $\sim5\times10^{60}$~ergs. 
Thus the conclusion by \cite{soer17} that their simulation indicates that quasar feedback (of order $10^{60}$~ergs) changes the measured thermal energy by only 20 per cent is a statement about the thermal energy in the \textit{Planck} beam rather than that in the quasar host halo. 

Because ACT has nearly five times the resolution of \textit{Planck},  C16 and this study more directly probe the thermal energy of the quasar host itself, i.e., we expect less contamination from clustered dark matter halos. 
At 148~GHz the ACT beam size is approximately two times larger than the angular extent of the dark matter halos hosting  quasars at z$\sim$2 \citep{hall18}. While further analysis is needed to more precisely specify the SZ contribution from correlated structure, our results provide a new estimate of the quasar-correlated SZ with better filtering of the 2-halo term. Interpreted in this way, we have provided a new upper limit to the thermal energy in quasar host halos and to the feedback efficiency of quasars.

\section{Summary and Conclusions}

We construct model SEDs from the radio to far-infrared of 109,829 optically-selected quasars from the SDSS, divided into six redshift bins. 
We find that the SEDs are well fit with an optically thin synchrotron model in the radio, a two temperature modified blackbody dust emission model to describe the far-infrared emission, CO emission lines to describe excess emission at observed 95~GHz and 148~GHz at redshifts $z<1.91$, plus an additional spectral distortion from the thermal SZ effect. 
We investigate other sources for the excess millimeter emission, finding that free-free without CO and anomalous microwave emission cannot explain the excess. 
The addition of a free-free component while still including CO emission cannot be ruled out, but offers no improvement over CO alone. An optically thick synchrotron component (\citealt{ozel00,inou14,beha15}) with or without CO emission is the most viable alternative to explain the excess. 
The derived infrared emission is consistent with previous quasar studies with ${L_{\rm IR,tot} \sim 10^{12}}$~L$_\odot$ and effective dust temperatures in excess of that expected for non-active, star forming galaxies (\citealt{bian19, hall18, kirk12}). 
However, more work is needed to disentangle the quasar host emission from dusty star forming galaxies in the same halo.
The inclusion of thermal SZ effect in the baseline model is preferred at the $\sim$4$\sigma$ level when fitting all six redshift bins simultaneously.

Due to the uncertainty associated with understanding the excess emission at 95 and 148~GHz, we proceed with the interpretation of the amplitude of the thermal SZ signal in the three highest redshift bins at $z>1.91$.
In these bins, we fit a model using the synchrotron and two temperature modified blackbody emission models plus the spectral distortion from the thermal SZ effect. 
The SZ is preferred in the model at the level of 3.8$\sigma$.
Throughout this work, we parameterize the amplitude of the SZ effect in terms of the fraction of the bolometric luminosities of the quasars that thermally couples to the surrounding medium over a fiducial lifetime of 10$^8$ years.
Assuming the entire SZ signal is due to hot bubbles from quasar winds, we find that the total thermal energy measured amounts to 5$\tau_8^{-1}$ per cent of the quasar bolometric luminosities over this quasar lifetime. 
Given the range of models we tested (Table ~\ref{compare}), a coupling fraction of $f = 3-7 \tau_8^{-1}$ per cent is consistent with our data.
This range of values is the same as those used in simulations of galaxy evolution that reproduce the luminosity function of the observed universe \citep[e.g.,][]{hopk06}.
Translated to thermal energy, this level of SZ corresponds to $\sim3.1\times10^{60}$~ergs at $z\sim2$.
If the total thermal energy is due to virialization of the dark matter halo, this implies an average halo mass of $M_h=6.3\times10^{12} ~h^{-1}$M$_\odot$, which is $\sim1.5$ times more massive than recent measurements for radio quiet quasars (\citealt{Efte19, geac19}).
If instead the quasars are hosted by $M_h=2\times10^{12}~h^{-1}$M$_\odot$ dark matter halos, then the measured total thermal energy is fifteen times greater than that expected due to virialization.  Further work is needed to separate the thermal energy contribution of quasar feedback from that of virialization and correlated structure \citep{cen15,hill18}.

This work provides a new understanding of the emission mechanisms in the environment of radio quiet quasars and how the parameters describing this emission affect the quasar-correlated thermal SZ measurement. 
A breadth of observations at millimeter wavelengths in the form of large surveys (such as from Advanced ACTPol, the South Pole Telescope 3G, and the future Simons Observatory and CMB-S4) and targeted observations of individual objects will illuminate the emission mechanisms and physical scales on which they dominate. 
The SZ effect is a promising way forward to understanding the total thermal energy of the bulk of the gas surrounding quasars, and future observations are needed in order to fully understand the coupling of the quasar wind radiation to the surrounding medium.
Higher resolution data at millimeter wavelengths is needed to better understand the emission of radio-quiet quasars in this regime, to discover the quasar-driven hot bubbles, and to disentangle contributions from the virialized halo gas and quasar feedback in individual systems.

\section*{Acknowledgements}

We thank the anonymous referee for their careful review and improvement of this manuscript. 

This work was supported by the U.S. National Science Foundation through awards AST-1440226, AST-0965625 and AST-0408698 for the ACT project, as well as awards PHY-1214379 and PHY-0855887. Funding was also provided
by Princeton University, the University of Pennsylvania,
and a Canada Foundation for Innovation (CFI) award
to UBC. ACT operates in the Parque Astronomico Atacama in northern Chile under the auspices of the Comision Nacional de Investigacion Cientfica y Tecnologica
de Chile (CONICYT).

The \textit{Herschel}-ATLAS is a project with \textit{Herschel}, which
is an ESA space observatory with science instruments
provided by European-led Principal Investigator consortia and with important participation from NASA. The
H-ATLAS website is http://www.h-atlas.org/.

Part of this research project was conducted using computational resources at the Maryland Advanced Research Computing Center (MARCC).

DC acknowledges the financial assistance of the South African Radio Astronomy Observatory (SARAO, www.ska.ac.za).
KM acknowledges support from the National Research Foundation of South Africa. LM is funded by CONICYT FONDECYT grant 3170846.

\bibliographystyle{mnras}
\clubpenalty=10000
\widowpenalty=10000
\interlinepenalty=10000
\bibliography{kh_master}

\bsp	
\label{lastpage}
\end{document}